\PassOptionsToPackage{pdfpagelabels=false}{hyperref} 
\documentclass[fleqn,usenatbib]{mnras}

\usepackage{newtxtext,newtxmath}

\usepackage[T1]{fontenc}
\usepackage{ae,aecompl}

\usepackage{graphicx}	
\usepackage{amsmath}	
\usepackage{amssymb}	
\usepackage{mathrsfs}
\usepackage{soul,xcolor}

\def\apj{Astrophys. J.}
\def\apjl{Astrophys. J. Lett.}
\def\apjs{Astrophys. J. Supp. Ser. }

\def\aap{Astron. Astrophys. }

\def\physrep{Phys. Rep. }
\def\mnras{Mon. Not. Roy. Astron. Soc. }

\def\apss{Astrophys. Space Sci.}

\title[Vorticity waves and shock dynamics in CCSNe]{The impact of vorticity waves on the shock dynamics in core-collapse supernovae}

\author[C. Huete, E. Abdikamalov, and D. Radice]{
C\'esar Huete$^{1}$\thanks{E-mail: chuete@ing.uc3m.es},
Ernazar Abdikamalov$^{2}$\thanks{E-mail: ernazar.abdikamalov@nu.edu.kz}, and David Radice$^{3,4}$\thanks{E-mail: dradice@astro.princeton.edu}
\\
$^{1}$ Fluid Mechanics Group, Escuela Polit\'ecnica Superior, Universidad Carlos III de Madrid, SP\\
$^{2}$ Department of Physics, School of Science and Technology, Nazarbayev University, Astana 010000, KA \\
$^{3}$ Institute for Advanced Study, 1 Einstein Drive, Princeton, NJ 08540, USA \\
$^{4}$ Department of Astrophysical Sciences, Princeton University, 4 Ivy Lane, Princeton, NJ 08544, USA
}

\date{Accepted XXX. Received YYY; in original form ZZZ}

\pubyear{2017}

\begin{document}
\label{firstpage}
\pagerange{\pageref{firstpage}--\pageref{lastpage}}
\maketitle

\begin{abstract}
Convective perturbations arising from nuclear shell burning can play an important role in propelling neutrino-driven core-collapse supernova explosions. In this work, we analyze the impact of vorticity waves on the shock dynamics and the post-shock flow using the solution of the linear hydrodynamics equations. We show that the entropy perturbations generated by the interaction of the shock with vorticity waves may play a dominant role in generating buoyancy-driven turbulence in the gain region. We estimate that the resulting reduction in the critical luminosity is $\sim 17-24\%$, which approximately agrees with the results of three-dimensional neutrino-hydrodynamics simulations.  
\end{abstract}





\section{Introduction}

First described by \citet{Baade34} as the "transition of an ordinary star into a neutron star", core-collapse supernovae (CCSNe) are the powerful explosions of massive stars that occur at the end of their lives \citep[e.g.,][]{Bethe90}. Upon reaching its maximum mass, the iron core becomes unstable, initiating a collapse to a proto-neutron star (PNS). The shock wave launched at core bounce quickly loses its energy and stalls at a radius of $\sim 150\,\mathrm{km}$. In order to produce an explosion and leave behind a stable neutron star, the shock must recover within a few hundreds of milliseconds and expel the stellar envelope \citep[e.g.,][]{Oconnor11,Ott11,Ugliano12}. Otherwise, a black hole (BH) forms \citep[e.g.,][]{Nadezhin80,Lovegrove13,Adams17,Kashiyama15}. The details of how this occurs remain unclear, constituting one of the longest-standing open questions in modern astrophysics \citep[see, e.g.,][for recent reviews]{Janka12, Foglizzo15, Burrows13, Mueller16b}. 

A key ingredient for producing the explosion is the neutrino emission by the newly-born PNS, which deposits energy behind the shock and establishes a negative entropy gradient that drives vigorous neutrino-driven convection. Together with the standing-accretion shock instability (SASI), these multi-dimensional hydrodynamic effects create favorable conditions for shock revival \citep[e.g.,][]{Herant95, Burrows95, Janka96, Blondin03, Foglizzo06, Yamasaki06,Hanke12, Hanke13,  Dolence13, Murphy13, Takiwaki14, Ott13, Abdikamalov15, Radice15, Radice16, Melson15a, Lentz15, Fernandez14, Fernandez15, Cardall15, Bruenn16, Roberts16}. If present, rapid rotation may facilitate explosion via the magneto-rotational mechanism \citep{Burrows07,Moesta14,Moesta15} \citep[see also][]{Takiwaki16,Summa17}. 

\citet{Couch13} demonstrated that the perturbations arising from the turbulent convection in Si and O burning shells in CCSN progenitors may help to revive the shock. As the iron core collapses, the perturbations follow the core and accrete towards the center. Due to the converging geometry of the flow, the perturbations amplify significantly during collapse \citep{Kovalenko98,Lai00,Takahashi14}. Further amplification occurs at shock crossing \citep{Abdikamalov16}. Once in the post-shock region, the fluctuations contribute to the non-radial flow in the gain region, creating a more favorable condition for producing explosion \citep{Couch15a, Couch15b, Mueller15, Abdikamalov16, Takahashi16, Burrows16, Radice17}. 

\citet{Mueller16} presented 3D simulation of the last minutes of O shell burning in an $18M_\odot$ progenitor star. Prior to collapse, they observed vigorous convection with Mach number of $\sim 0.1$ and dominant angular wave number of $l=2$. Full 3D neutrino-hydrodynamics simulation of this model yielded strong explosion after the accretion of the O shell through the shock, whereas in a model with artificially suppressed pre-collapse convection, no explosion was observed \citep{Mueller17}. The reduction of the the critical (i.e., minimum) neutrino luminosity for producing explosion due to these perturbations was estimated to be $\sim 20\%$, which is roughly in agreement with the analytic predictions of \citet{Mueller16}. Recently, \citet{Collins17} investigated the properties of Si and O shell burning in a broad range of presupernova models with ZAMS masses between $9.45M_\odot$ and $35M_\odot$. They found that the progenitor models between $16M_\odot$ and $26M_\odot$ exhibit large scale convective motions with high Mach numbers in the O shells, which are favorable conditions for producing perturbation-aided neutrino-driven explosions \citep{Mueller15}. On the other hand, strong perturbations were rarely observed in the Si shells. 

The emerging qualitative picture of how the progenitor asphericities impact the explosion condition is as follows. The convective vorticity waves distort the spherical isodensity surfaces of the progenitor star, creating Eulerian density perturbations at a given radius. When these density and vorticity perturbations encounter and cross the shock, they generate strong buoyancy-driven turbulence in the post-shock region, which helps to trigger an explosion \citep{Mueller15}.  

In order to gain the full understanding of how these perturbations affect the explosion dynamics, it is necessary to understand the physics of shock-turbulence interaction, starting with linear order. With this premise, \citet{Abdikamalov16} studied the effect of entropy and vorticity perturbations using a linear perturbation theory known as the {\it linear interaction analysis} (LIA) \citep[e.g.,][]{Ribner53,Mahesh96}. These two represent two of the three components of a generic turbulent flow, the third being acoustic waves \citep{Kovasznay53,Chu1958}. They found that the kinetic energy of these fluctuations increases by a factor of $\sim 2$ as they cross the shock. Assuming direct injection of this energy into the post-shock region, they estimated that these perturbations can reduce the critical neutrino luminosity for producing explosion by $\sim 12\%$. While this is an important finding, the physics of shock-turbulence in CCSNe, even at linear level, is not yet completely understood. As noted by \citet{Mueller17}, the buoyancy plays a dominant role in generating post-shock turbulence. Moreover, the acoustic waves generated by infalling entropy and vorticity perturbations \citep{Kovalenko98, Foglizzo00, Mueller16b} will affect the shock dynamics and the post-shock flow. Finally, the impact of perturbations on the nuclear dissociation rate itself should also be taken into account. These aspects are missing from the analysis of \citet{Abdikamalov16}. 

In this work, we investigate the interaction between accretion shocks and turbulent fluctuations in further detail. Our study is based on the solution of the linearized hydrodynamics equations in the post-shock region, which permits to capture the full temporal evolution of shock-vorticity interaction. The mathematical formalism describing the post-shock perturbation flow is similar to that employed in theoretical works on Richtmyer-Meshkov-type flows \citep{Wouchuk2001,Wouchuk2001b,Cobos2014} and analogous to that used in canonical interactions of non-reactive and reactive shocks with turbulent flows \citep{Wouchuk2009,Huete2017}. This improved formalism allows us to take into account the perturbation of the nuclear dissociation itself, which was not included in \citet{Abdikamalov16}. As demonstrated below, this effect is found to be important in the turbulent kinetic amplification factor, with the parametric trends and the asymptotic values being significantly affected. 

This is the first in a series of two papers. The current paper is dedicated to the study of the interactions of accretion shocks with vorticity waves, while the second will study the interactions with density perturbations generated due to differential infall. The aim of this series of works is to establish in detail the linear physics of interaction of shocks with hydrodynamic turbulence in CCSNe. 

The rest of the paper is organized as follows. Section~\ref{sec:problem} presents the problem formulation and the solution method. In Section~\ref{sec:analysis_wave}, an analysis of the interaction of shock waves with individual vorticity waves is presented, while Section~\ref{sec:analysis_field} focuses on the interaction of shocks with isotropic field of vorticity waves. The base-flow properties for the shock Mach number and the dissociation degree are computed in Section~\ref{sec:varepsilon}. In Section~\ref{sec:discussion}, we discuss the implication of our results on the explosion condition of core-collapse supernovae. Finally, in Section~\ref{sec:conclusion} we present our conclusions.

\section{Problem Formulation}
\label{sec:problem}
\subsection{Perturbation-free flow}
\begin{figure}
\includegraphics[width=0.47\textwidth]{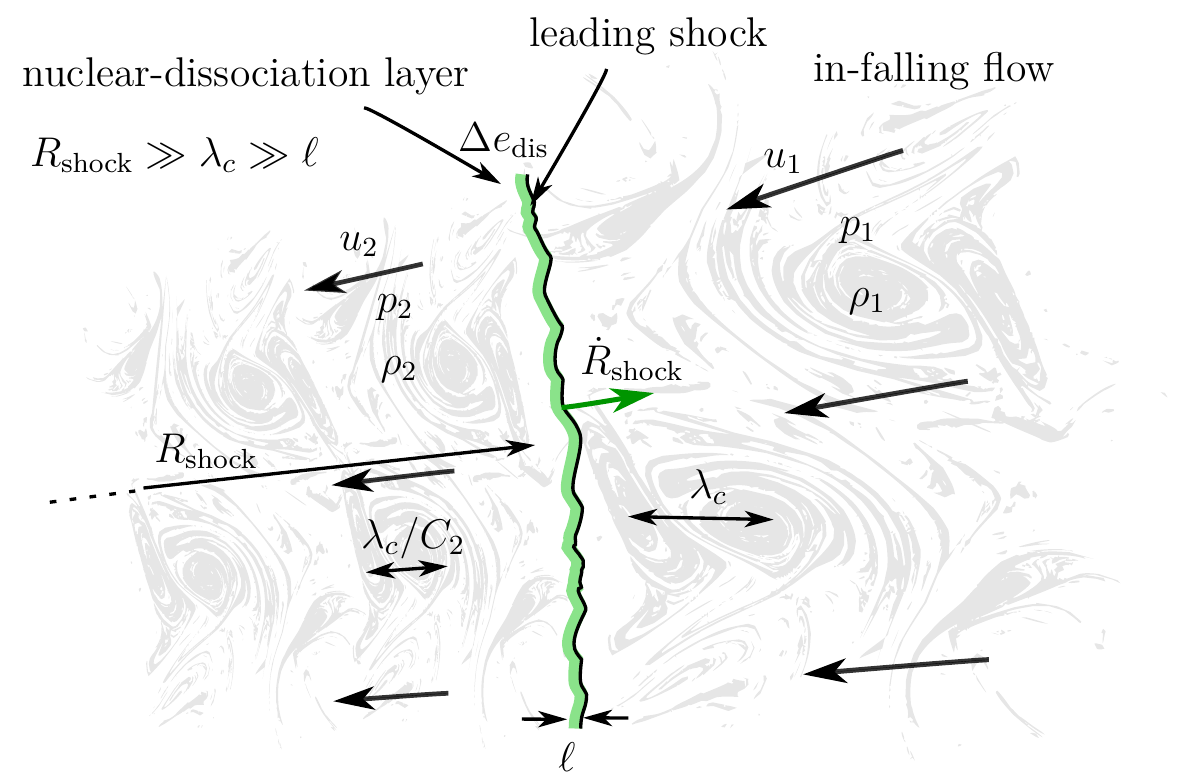}
\caption{Scheme of the accretion shock expanding through the in-falling mass and characteristic scales: shock radius $R_{{\rm shock}}$, shock thickness $l$, and characteristic perturbation wavelength $\lambda_c$.}
\label{fig:scheme2D}
\end{figure}

Let us consider an expanding shock wave placed at $r=R_{{\rm shock}}(t)$ that separates the in-falling flow ahead of shock front $r>R_{{\rm shock}}$, denoted with subscript 1, and the downstream post-shock flow identified with subscript 2 in $r<R_{{\rm shock}}$ (see Fig. \ref{fig:scheme2D} for clarification). In the thin-shock limit, when the radius of the shock is much larger than the accretion-shock thickness $R_{{\rm shock}}\gg l$, the variation of the different flow variables across the shock is readily obtained through the radial integration of the conservation equations, yielding
\begin{subequations}
\begin{alignat}{3}
&\rho_1 \left(u_1+\dot{R}_{{\rm shock}}\right) = \rho_2 u_2 \ ,\label{mass0}\\
&p_1 + \rho_1 \left(u_1+\dot{R}_{{\rm shock}}\right)^2 = p_2 + \rho_2 u_2^2\ ,\label{momentum0}\\
&e_1 +\frac{p_1}{\rho_1} +\frac{1}{2} \left(u_1+\dot{R}_{{\rm shock}}\right)^2 = e_2 +\frac{p_2}{\rho_2} + \frac{1}{2} u_2^2\ ,\label{energy0}%
\end{alignat}
\end{subequations}
for the mass, momentum and energy conservations equations, respectively. The symbols $u$, $\rho$, $p$ and $e$ refer to the bulk velocity, density, pressure, and internal energy of the gas, respectively. Notice that, for non-negligible accretion shock thicknesses, the mass equation \eqref{mass0} should include the term involving the divergence of the post-shock expanding gas.

When the compressed flow, modeled as a perfect gas with the polytropic index $\gamma=4/3$, is affected by nuclear dissociation effects occurring in a thin layer right behind the shock, the variation of the internal energy can be computed as 
\begin{equation}
e_1-e_2 =\frac{1}{\gamma-1}\frac{p_1}{\rho_1}-\frac{1}{\gamma-1}\frac{p_2}{\rho_2} + \Delta e_\mathrm{dis}
\end{equation}
with $\gamma$ assumed constant through the interaction process, and $\Delta e_\mathrm{dis}$ referring to the energy per unit mass employed in dissociating the nuclei. 

Following \citet{Fernandez2009a,Fernandez2009b}, the nuclear dissociation energy can be scaled with the free-fall speed squared,
\begin{equation}
\Delta e_\mathrm{dis} = \frac{1}{2} \varepsilon \upsilon_\mathrm{FF}^2,
\label{Deltaedis}
\end{equation}
using dimensionless nuclear dissociation parameter $\varepsilon$. As we show below in Section~\ref{sec:varepsilon}, $\varepsilon$ typically ranges from $0.2$ to $0.4$ in CCSN models. Assuming that the Bernoulli parameter is zero above the shock,  
\begin{equation}
\upsilon_\mathrm{FF}^2 = \frac{2 G M}{R_{{\rm shock}}} = \frac{1}{2} u_1^2 +  \frac{\gamma}{\gamma-1}\frac{p_1}{\rho_1}\ ,
\end{equation}
where $G$ is the gravitational constant and $M$ is the gravitating mass. 

In the stalled-shock regime, $u_1/\dot{R}_{{\rm shock}}\gg1$, the Mach number $M_1 =u_1/a_1$, with $a_1=(\gamma_1 p_1/\rho_1)^{1/2}$ defining the speed of sound upstream, is used to rewrite nuclear dissociation energy as
\begin{equation}
\frac{\gamma^2-1}{2}\frac{\Delta e_\mathrm{dis}}{a_1^2} = \varepsilon  \frac{\gamma+1}{2}\left(1+\frac{\gamma-1}{2}M_1^2\right).
\label{eq:varepsilon}
\end{equation}

Taking $\varepsilon$ and Mach as the independent parameters, the values of which may vary within the range established from numerical simulations of CCSNe (see Section~\ref{sec:varepsilon}), the fluid properties behind the shock are expressed in the form
\begin{equation}
C_2 = \frac{\rho_2}{\rho_1} = \frac{u_1}{u_2} = \frac{\left( \gamma + 1\right) M_1^2}{\left( \gamma - \kappa  \right) M_1^2 + 1}\ ,
\label{R}
\end{equation}
and
\begin{equation}
P_2 =\frac{p_2}{\rho_1 u_1^2}=\frac{\gamma M_1^2(1+\kappa) +1}{\gamma(\gamma + 1) M_1^2}\ ,
\label{P}
\end{equation}
for post-shock density and pressure. The Mach number of the fluid particles leaving the shock is 
\begin{equation}
M_2 = \frac{u_2}{a_2} = \left(\gamma C_2 P_2\right)^{-1/2} =  \left[\frac{\left( \gamma - \kappa  \right) M_1^2 + 1}{\gamma M_1^2(1+\kappa) +1}\right]^{1/2}\ ,
\label{M2}
\end{equation}
with the function
\begin{equation}
\kappa=\left[(1-M_1^{-2})^2+ \varepsilon(\gamma+1) \left(\gamma-1+2 M_1^{-2}\right)\right]^{1/2}
\label{kappa}
\end{equation}
accounting for the dimensionless endothermic parameter $\varepsilon$. For non-reacting shock waves the value of $\kappa=1-M_1^{-2}$, thereby reducing Eqs.~\eqref{R}-\eqref{M2} to the well-known regular Rankine-Hugoniot relationships.

\begin{figure}
\includegraphics[width=0.5\textwidth]{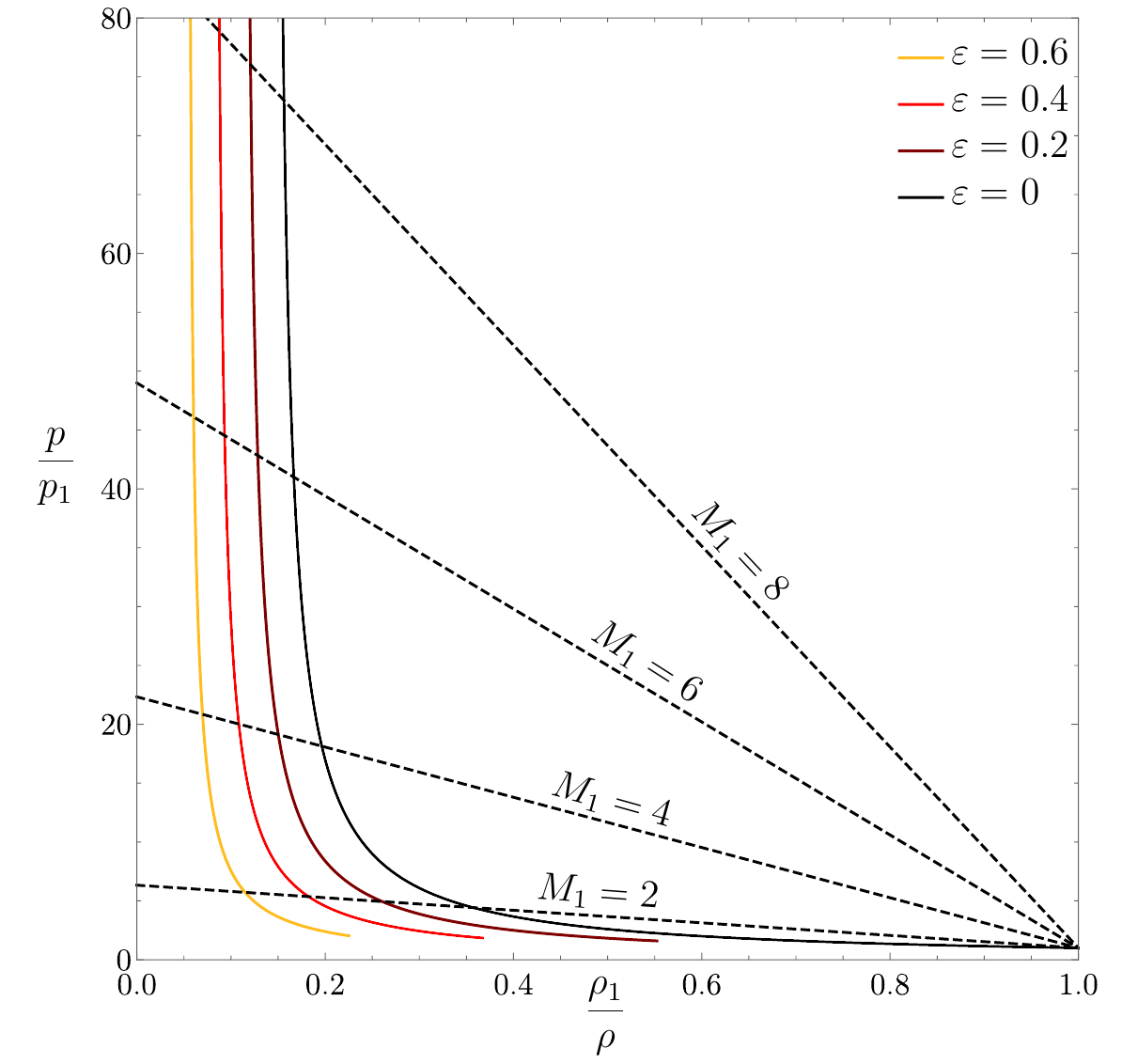}
\caption{Hugoniot curves and Rayleigh lines for several values of the dissociation energy $\varepsilon = 0,\ 0.2,\ 0.4$ and $0.6$.}
\label{fig:RH}
\end{figure}

The effect of nuclear dissociation into the post-shock flow density and pressure is easily analyzed through Fig. \ref{fig:RH}, with the final values provided by the intersection of the Rayleigh line (for a constant Mach number propagation) and the non-adiabatic Rankine-Hugoniot curve. It is found that higher pressures and densities are required downstream to get the shock with the same Mach number, if endothermic transformations take place through the shock wave. The maximum energy that can be employed in the nuclear dissociation process is found in the extreme limit $1-\varepsilon\ll1$, which provides limiting conditions for the post-shock gas: post-shock Mach number and temperature that tends to zero, and density that tends to infinity. That is, all the kinetic and thermal energy of the in-falling gas is used in dissociating the nuclei. The corresponding Hugoniot curve collapses into the vertical axis and the finite values of pressure are given by the intersection with the Rayleigh lines.

\begin{figure}
\includegraphics[width=0.47\textwidth]{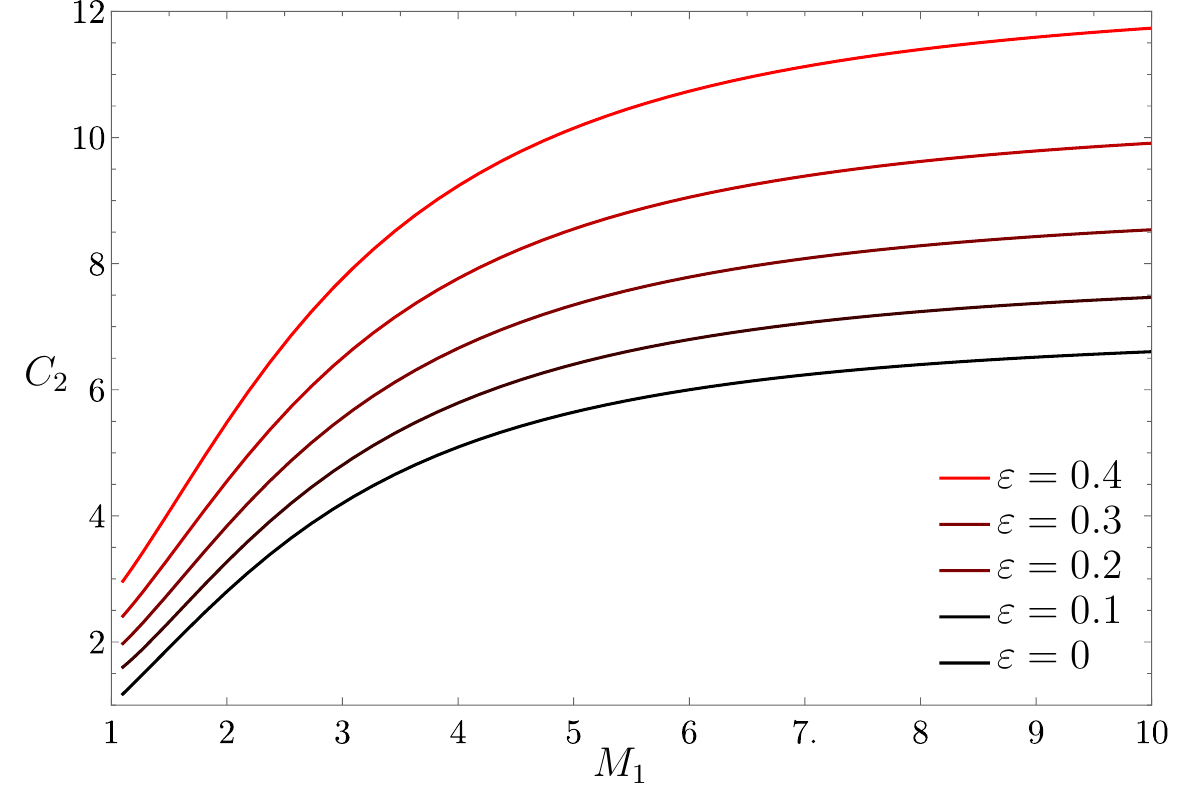}\\
\includegraphics[width=0.47\textwidth]{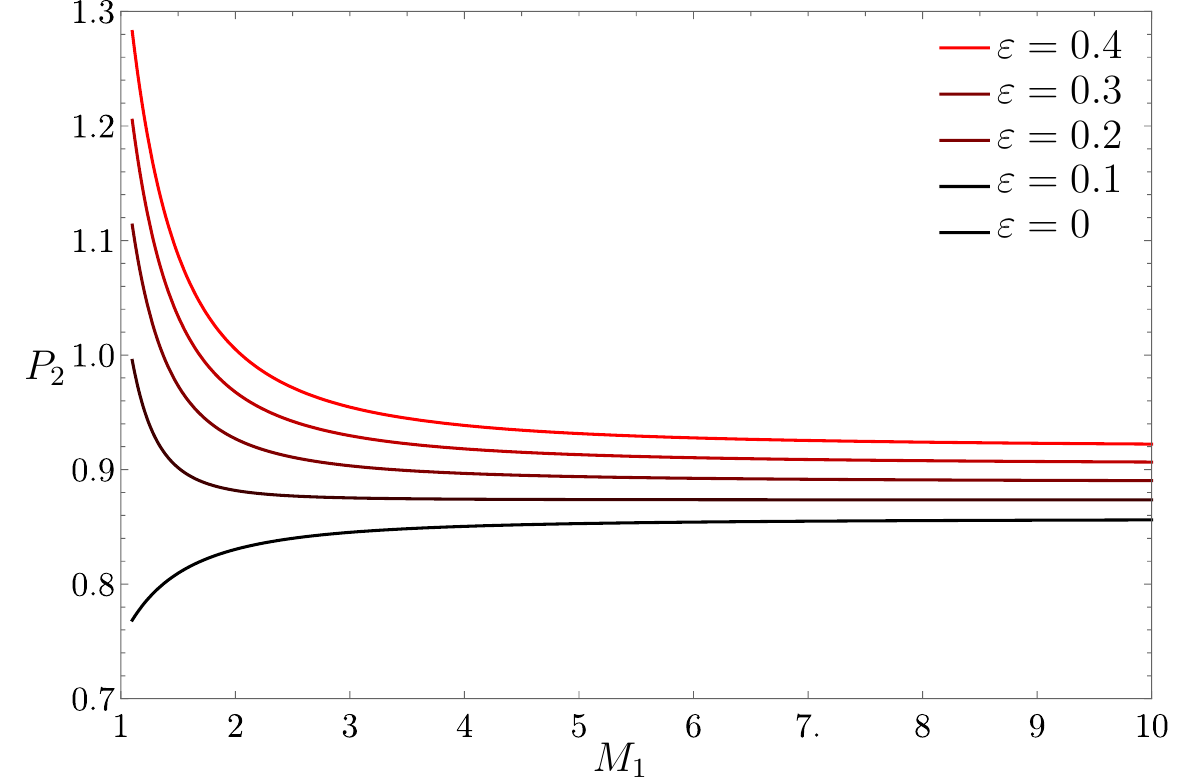}
\caption{Mass compression ratio $C_2$ (Top) and pressure amplification $P_2$ (Bottom) as a function of the incident Mach number $M_1$ for $\varepsilon = 0,\ 0.1,\ 0.2,\ 0.3$ and $0.4$}
\label{fig:RP}
\end{figure}

Equations \eqref{R} and \eqref{P} are computed in Fig. \ref{fig:RP} as a function of the Mach number, $M_1$. Both $C_2$ and $P_2$ increase with $\varepsilon$ for the same value of $M_1$. Unlike regular detonations, where chemical energy release does not depend on shock intensity since the reaction is self-sustaining, nuclear-dissociation degree does depend on upstream Mach number. Thus, the function $\kappa$ approaches the value $\left[1+(\gamma^2-1)\varepsilon\right]^{1/2}$ in the strong-shock limit, $M_1\gg1$, then yielding 
\begin{equation}
C_2|_{M_1\gg1} = \frac{\gamma + 1}{ \gamma - \left[1+\left(\gamma^2-1\right)\varepsilon\right]^{1/2}}
\label{RMgg1}
\end{equation}
and
\begin{equation}
P_2|_{M_1\gg1} = \frac{1+ \left[1+\left(\gamma^2-1\right)\varepsilon\right]^{1/2}}{\gamma +1}\ ,
\label{PMgg1}
\end{equation}
for the post-shock density and pressure values, in agreement with Fig. \ref{fig:RP}. The mass-compression ration $C_2$ is found to diverge, and $P_2$ approaches unity in the double limit $M_1\gg1$, $1-\varepsilon\ll1$.

\subsection{Perturbation problem}
\label{sec:perturb_problem}
The upstream and the downstream linear disturbances can be characterized in terms of acoustic, entropy and vortical modes. In the in-falling gas reference frame, the upstream mono-frequency perturbation in the fresh-gas reference frame $(x_1,y_1)$ is determined by the divergence-free velocity perturbation wave, namely
\begin{equation}
\begin{aligned}
&\bar{u}_1 \left( x_1, y_1 \right)= \frac{u_1-\langle u_1 \rangle}{ \langle a_2 \rangle}
= \hat{u}_1 \cos\left(k_x x_1 \right)\cos\left(k_y y_1 \right), \\
&\bar{v}_1 \left( x_1, y_1 \right)= \frac{v_1-\langle v_1 \rangle}{ \langle a_2 \rangle}
=  \hat{u}_1  \frac{k_x}{k_y} \sin\left(k_x x_1 \right)\sin\left(k_y y_1 \right),
\label{u1v1}%
\end{aligned}
\end{equation}
for the streamwise and crosswise perturbations, respectively. The brackets denote the time-averaged mean value of the flow variable, which is effectively null for the upstream velocity in the stagnant gas reference frame. The dimensionless factor $\hat{u}_1$ stands for the amplitude of the upstream velocity disturbances and $\vec{k}=(k_x,k_y)$ is the upstream wave number vector. The associated non-dimensional vorticity wave, $\bar{\omega}_1 \left( x_1, y_1 \right) =\partial \bar{v}_1 /(\partial k_y x_1) -\partial \bar{u}_1/(\partial k_y y_1)$, is
\begin{equation}
\bar{\omega}_1 \left( x_1, y_1 \right) = \hat{u}_1 \left(1+\frac{k_x^2}{k_y^2}\right)  \cos\left(k_x x_1 \right)\sin\left(k_y y_1 \right)\ .
\label{omega1}
\end{equation}

\begin{figure}
\includegraphics[width=0.5\textwidth]{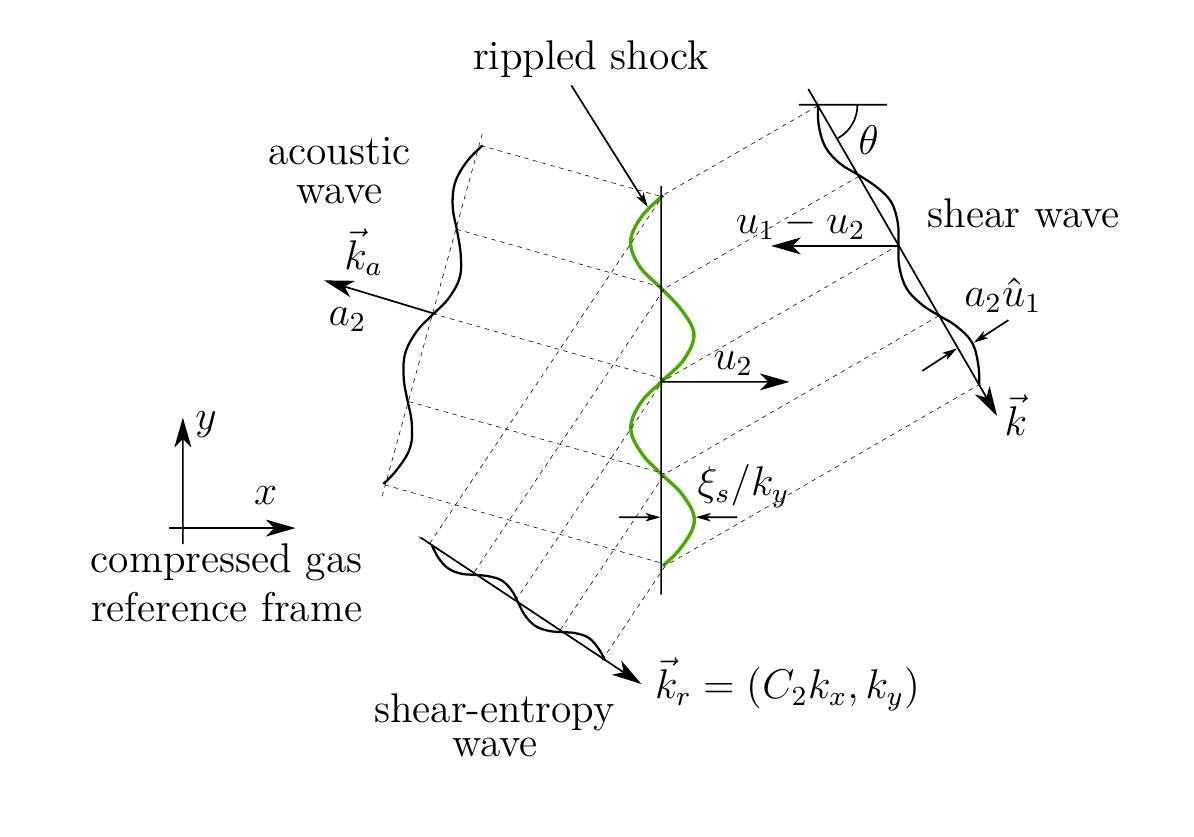}
\caption{Scheme of the shock shear-wave interaction in the compressed gas reference frame.}
\label{fig:scheme}
\end{figure}

The interaction of the CCSN shock with the upstream shear wave, characterized by the angle $\theta = \tan^{-1}(k_y/k_x)$, is sketched in Fig. \ref{fig:scheme}. As a result of the interaction, the shock ripples and the fluid downstream is correspondingly altered with acoustic and entropic-vorticity waves, the former traveling at the speed of sound downstream $a_2$ and the latter moving with the fluid particles.

For the perturbed accretion shock to be seen as a discontinuity front, the characteristic perturbation wavelength $\lambda_c\sim k_y^{-1}$ must be much larger than the accretion-shock thickness $l$, including the dissociation layer in it (see Fig. \ref{fig:scheme2D}). Besides, to consider the base-flow variable as constant properties, the shock and the in-falling gas must be in the nearly-steady regime, so that the variations of the base-flow properties within the characteristic wavelength can be neglected. These two conditions are simultaneously true for perturbation wavelengths much smaller than expanding shock evolution range and much higher than the shock thickness, both used to define the limits of validity of the model associated to spatial and temporal scales: $k_y R_{{\rm shock}} \gg 1\gg k_y l$, and $ \dot{R}_{{\rm shock}}\ll a_2 \dot{\xi}_s$, respectively. On the other hand, the planar shock assumption, $k_y R_{{\rm shock}} \gg 1$, is not suitable for perturbations characterized by low-mode numbers such as SASI \citep{Blondin03,Fernandez15}. For such modes, spherical geometry is more suitable \citep{Foglizzo09}, which has been employed by \citet{Takahashi16} to study the influence of pre-collapse perturbations on the hydrodynamic eigenmodes in the gain region. 

For the analysis it is convenient to use a reference frame moving with the velocity of the post-shock flow. The solution is to be described in terms of the dimensionless coordinates $x = k_y x_2$ and $y = k_y y_2$ and the dimensionless time $\tau = a_2 k_y t $. 

The nondimensional values for pressure, density and velocity perturbations downstream, defined as
\begin{equation}
\begin{aligned}
&\bar{p} = \frac{p-\langle p_2 \rangle}{\gamma \langle p_2 \rangle} \ ,\quad & &\bar{\rho}= \frac{\rho-\langle \rho_2 \rangle}{\langle \rho_2 \rangle}\ , \\
&\bar{u}=  \frac{u-\langle u_2 \rangle}{\langle a_2 \rangle} \ ,\quad &
&\bar{v}= \frac{v-\langle v_2 \rangle}{\langle a_2 \rangle}\ ,
\label{perturb}
\end{aligned}
\end{equation}
respectively, are used to write the adiabatic Euler equations governing the post-shock flow. Anticipating that $\bar{p}$ and $\bar{v}$ are always proportional to $\cos(y)$ and $\sin(y)$, respectively, the conservation equations for mass, $x$-momentum, $y$-momentum and energy, namely
\begin{equation}
\begin{aligned}
&\frac{\partial \bar{\rho}}{\partial \tau} +\frac{\partial \bar{u}}{\partial x}+\bar{v}=0 \ ,\quad &
&\frac{\partial \bar{u}}{\partial \tau} +\frac{\partial \bar{p}}{\partial x}=0\ ,\\
&\frac{\partial \bar{v}}{\partial \tau} -\bar{p}=0 \ ,\quad &
&\frac{\partial \bar{p}}{\partial \tau} = \frac{\partial \bar{\rho}}{\partial \tau}\ ,
\label{eqperturb}
\end{aligned}
\end{equation}
respectively, are combined for $\bar{p}$ to yield
\begin{equation}
\frac{\partial^2 \bar{p}}{\partial \tau^2}= \frac{\partial^2 \bar{p}}{\partial x^2}-\bar{p}
\label{sonicwave}
\end{equation}
as the two-dimensional periodically-symmetric wave equation, which governs the perturbation field behind the shock.

The problem reduces to that of integrating the linearized Euler equations, or equivalently the wave equation \eqref{sonicwave}, for $\tau\geq0$ and within the domain delimited by the leading reflected sonic wave traveling backwards, $x= -\tau$ and the shock front moving upwards $x= M_2\tau$. One boundary condition is provided by the isolated-shock assumption, which translates into not considering the effect of the acoustic waves reaching the shock front from behind, in consonance with the large radius limit, $R_{{\rm shock}} k_y \gg 1$. On the other hand, the boundary condition at the CCSN shock is determined by the linearized Rankine-Hugoniot relationships,
\begin{subequations}
\begin{alignat}{3}
&\left(C_2-1\right)\dot{\xi}_s =C_2\bar{u}_s-M_2 C_2\bar{\rho}_s-\bar{u}_1 \ , \label{massRH}\frac{}{}\\
&\bar{p}_s = 2M_2\left(\bar{u}_s-\bar{u}_1 \right)-M_2^2\bar{\rho}_s\ , \label{xmomRH}\frac{}{}\\
&M_1^2 M_2^2\bar{\rho}_s = \Pi_s\bar{p}_s-\Delta_s \left(\dot{\xi}_s-\bar{u}_1\right)\ , 
\label{eneRH}\frac{}{}\\
&\bar{v}_s = M_2 \left(C_2-1\right)\xi_s+\bar{v}_1\ ,\label{tanRH}
\end{alignat}
\end{subequations}
with $\dot{\xi}_s$ denoting the temporal derivative of the dimensionless ripple shock position $\xi_s = k_y\left( x_{1,s}-u_1 t\right)$, as depicted in Fig. \ref{fig:scheme}.

The energy equation \eqref{eneRH}, which involves the functions
\begin{equation}
\Pi_s = \frac{M_1^2\left[1 + M_1^2\left(1-\kappa\right)\right]^2}{\left(M_1^2+1\right)^2-M_1^4\kappa^2}
\label{Pis}
\end{equation}
and
\begin{equation}
\Delta_s = \varepsilon\frac{2 M_2^3 M_1^6\left(\gamma-1 \right) \left[1 + M_1^2\left(1-\kappa\right)\right]}{\left(M_1^2+1\right)^2-M_1^4\kappa^2}\ ,
\label{Deltas}
\end{equation}
distinguishes regular adiabatic shocks from reacting shocks like detonations or nuclear-dissociating shocks. 

In the previous work \citep{Abdikamalov16}, the coefficients accompanying the linear perturbations in the linear energy equation \eqref{eneRH} were the same as those found in perturbed adiabatic shock ($\Pi_s=1$ and $\Delta_s=0$), although the values of the base-flow properties, namely $M_2$ and $C_2$, were accordingly modified by nuclear dissociation effects. How nuclear-dissociation degree is affected by the perturbations, and how that modification ultimately acts upon the downstream flow variables, is incorporated in this model through coefficients $\Pi_s$ and $\Delta_s$. In this sense, the present analysis consistently accounts for the effect of $\varepsilon$ in both zero-order and first-order flow variables.

The value of $\Pi_s$ is positive when the dissociation energy is sufficiently low, that is $\kappa<1+M_1^{-2}$. On the other hand, when the dissociation energy is sufficiently high, the value of $\Pi_s$ becomes negative, then reverting the relationship between density and pressure perturbations in \eqref{eneRH}. Since the degree of dissociation depends on the shock strength, the term involving the function $\Delta_s$ in \eqref{eneRH} is proportional to the incident Mach number perturbation $\delta M_1 = \left(\dot{\xi}_s-\bar{u}_1 \right)M_1/(M_2 C_2)$. The value of $\Delta_s$ is found to be negative for $\varepsilon > 0$. It is worth commenting that the case of exothermic detonations is significantly different since the second term in the right-hand side of \eqref{eneRH} will vanish \citep{Huete2013,Huete2017}. This is so because the total heat release, generated by the combustion process behind the shock, does not depend on the shock intensity perturbation, as it is provided by self-sustained reactions. Once the reaction is triggered it will release all the thermonuclear (or chemical) energy.

Algebraical manipulation of \eqref{massRH}-\eqref{tanRH} is carried out to write one of the two equations for the shock boundary condition involving $\bar{\xi}_s$ and $\bar{p}_s$, that is
\begin{equation}
\frac{d \xi_s}{d \tau} = \sigma_a \bar{p}_s+\hat{u}_1\cos\left(\frac{k_x}{k_y}C_2 M_2 \tau\right)\ ,
\label{xis}
\end{equation}
with the  factor accompanying the pressure perturbation being
\begin{equation}
\sigma_a = \frac{C_2\left(M_1^2-\Pi_s\right)}{2 M_2 M_1^2  \left(C_2-1\right)+C_2\Delta_s}\ .
\label{As}
\end{equation}

Similarly, the material derivative behind the shock, $\partial/(\partial \tau) + M_2\partial/(\partial x)$, of the streamwise velocity perturbation $\bar{u}_s= \sigma_b \bar{p}_s+\bar{u}_1$, with
\begin{equation}
\sigma_b = \frac{M_1^2+\Pi_s+\Delta_s \sigma_a}{2 M_2 M_1^2}\ ,
\label{Bs}
\end{equation}
is used to provide
\begin{equation}
\begin{aligned}
&\left(\sigma_b +M_2 \right)\frac{\partial \bar{p}_s}{\partial \tau} +\left.\left(\sigma_b M_2+ 1\right)\frac{\partial \bar{p}}{\partial x}\right|_s=-M_2^2 \left(C_2- 1\right)\xi_s \\
&+\frac{k_x}{k_y}M_2\left(C_2-1\right)\hat{u}_1 \sin\left(\frac{k_x}{k_y}C_2 M_2 \tau\right)
\label{ps}
\end{aligned}
\end{equation}
as the second equation that conforms, along with \eqref{xis}, the shock boundary condition for the functions $\bar{\xi}_s$ and $\bar{p}_s$.

The coefficients $\sigma_a$ and $\sigma_b$ are positive for any combination of parameters $M_1$ and $\varepsilon$. In the strong shock limit for $\varepsilon>0$, the value of $\sigma_a$ approaches to zero with $\sigma_a|_{M_1\gg1}\sim M_1^{-2}$, while $\sigma_b$ reaches a constant value determined by the inverse of the post-shock Mach number, namely $\sigma_b|_{M_1\gg1}=M_2^{-1}|_{M_1\gg1}$.

The initial condition of the shock perturbations is determined by knowing that the shock is initially planar, so that $\bar{\xi}_s= \bar{v}_s = 0$. Correspondingly, the initial perturbations of pressure and streamwise velocity must satisfy $\bar{u}_s +\bar{p}_s=0$, as dictated by the first acoustic wave emitted backwards, thereby giving
\begin{equation}
\bar{p}_{s0}= -\frac{1}{\sigma_b+1} \hat{u}_1
\label{ps0}
\end{equation}
for the initial shock pressure perturbation.

\section{Linear interaction analysis with monochromatic vorticity perturbations}
\label{sec:analysis_wave}
\subsection{Shock pressure and corrugation temporal evolution}

The asymptotic behavior of the corrugated shock can be inferred from the Laplace transform expression provided in \eqref{PsLaplace}, with the imaginary poles in the dispersion relationship
\begin{equation}
\left(s\sqrt{s^2+1}+\sigma_b s^2 + \sigma_c\right)\left(s^2+\zeta^2\right)=0\ ,
\label{denominator}
\end{equation}
indicating the possibility of asymptotic harmonic oscillations. The left-hand product in \eqref{denominator} accounts for the shock response in the absence of continuous perturbations, whereas the right-hand product refers to the induced oscillations from the non-homogeneous upstream flow. The characteristic dimensionless frequency $\zeta$ is provided in \eqref{zeta}. Notice that the term $\sqrt{s^2+1}$ may change the sign if the pole lies on the bottom half-space of the imaginary plane.

It has been found that equation $s\sqrt{s^2+1}+\sigma_b s^2 + \sigma_c =0$ has no poles, indicating that shock pressure perturbations decay with time in the absence of continuous excitement. Generally, the perturbations decay in time like $\tau^{-3/2}$, but this decay rate changes, however, for infinitely strong shocks with $\varepsilon=0$, since $\sigma_b=\sigma_c$, yielding $\tau^{-1/2}$ as the law describing the approach to the permanent solution \citep{Fraley1986}.

We are firstly interested in the long-time response of the accretion shock to mono-frequency perturbations. As $\sigma_c<\sigma_b$ the shock will oscillate only with the excitement frequency coming from upstream perturbations, $\omega_s = R M_2 k_x/k_y$, thereby yielding an asymptotic response qualitatively similar to the one found for adiabatic shock waves \citep{Wouchuk2009}
\begin{equation}
\bar{p}_{s}(\tau\gg 1) =  \left \{ \begin{array}{ll}
\mathcal{P}_{lr} \cos\left(\omega_s \tau \right) + \mathcal{P}_{li} \sin\left( \omega_s \tau \right) & ,\zeta \leq 1 \\
\mathcal{P}_{s} \cos\left(\omega_s \tau \right) & ,\zeta \geq 1
	\end{array} \right.
\label{pstau}
\end{equation}
except for the coefficients defining the amplitudes, which are provided by Eqs.~\eqref{Plr}-\eqref{Ps} in Appendix~\ref{App1}. As the planar infinitely-thin assumption does not provide any length scale, the shock oscillation period will be proportional to the upstream characteristic length. In dimensional variables, the time between pressure peaks is given by $t_{\text{per}}=\lambda_x/(2 \pi a_1 M_1)$.

As in previous LIA works \citep{Wouchuk2009,Huete2013,Huete2017}, the pressure perturbation field splits into two distinguished regimes depending on the dimensionless frequency 
\begin{equation}
\zeta = \frac{k_x}{k_y}\frac{M_2 C_2}{\sqrt{1-M_2^2}}  = \frac{\omega_s}{\sqrt{1-M_2^2}}\ .
\label{zeta}
\end{equation}
In the long-wavelength (low-frequency) regime, $\zeta <1$, the acoustic perturbation right behind the shock is composed by the amplitudes of two orthogonal contributions $\mathcal{P}_{lr}$, and $\mathcal{P}_{li}$, respectively. In this range, the amplitude of the pressure disturbances exponentially decays with the distance from the shock front. On the other hand, in the short-wavelength (high-frequency) regime $\zeta >1$, the acoustic radiation travels in the form of constant-amplitude waves. The critical value $\zeta =1$ then indicates the condition at which stable sonic perturbations downstream move parallel to the shock front in the shock reference frame. 

As shown in equation~\eqref{Besselr} of Appendix~\ref{App1}, the post-shock pressure perturbation field can be computed as a linear combination of Bessel functions. In particular, right behind the shock, we have
\begin{equation}
\bar{p}_s(\tau)=\sum_{\nu=0}^{\infty}N_{\nu}J_{\nu}\left(r=\tau \sqrt{1-M_2^2}\right) \  ,
\label{psBesseltau}
\end{equation}
with the corresponding coefficients for $N_{\nu}$, provided in \eqref{Dm}, being obtained through the Laplace transform \eqref{Laplace} and the isolated-shock boundary condition. The temporal evolution of the shock ripple $\xi_s(\tau)$ is readily obtained through the integration of eq.~\eqref{xis}, whose solution can be expressed in terms of hyper-geometrical functions, as shown in \eqref{xisbeseel}. Akin to the shock pressure, the asymptotic long-time response is written in terms of harmonic functions, as provided in \eqref{xisasym}. 

\begin{figure}
\includegraphics[width=0.47\textwidth]{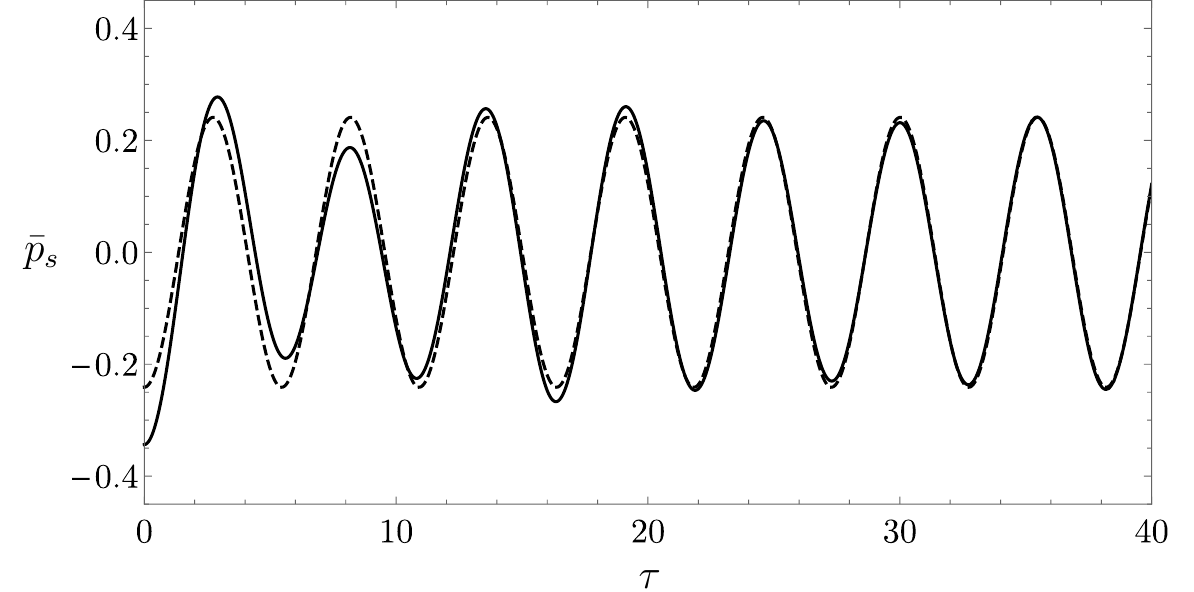} \vspace{2mm}
\\ \includegraphics[width=0.47\textwidth]{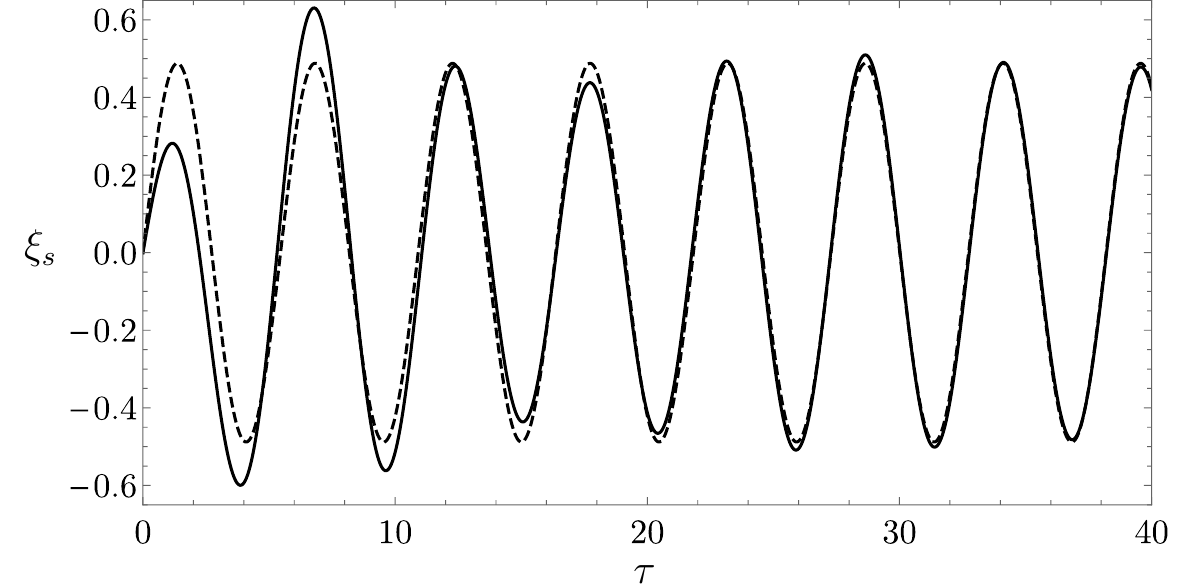} 
\caption{Shock pressure perturbation (top) and shock ripple amplitude (bottom) as a function of $\tau$ for $M_1=5$, $\zeta=1.2$, and for $\varepsilon = 0.4$. Solid: transient evolution \eqref{psBesseltau} and  \eqref{xisbeseel}. Dashed: asymptotic long-time equations \eqref{pstau} and \eqref{xisasym}.} 
\label{fig:pstau}
\end{figure}

The functions  $\bar{p}_s(\tau)$ and  $\xi_s(\tau)$ are computed in Fig. \ref{fig:pstau} as a function of $\tau$, for $M_1=5$, $\zeta=1.2$, and for $\varepsilon=0.4$. Both transient (solid line) and long-time response (dashed line) are shown. The shock transient evolution is found to agree fairly well with the asymptotic expressions provided in \eqref{pstau} and \eqref{xisasym}, then confirming that asymptotic functions can be used to compute the interaction with an isotropic spectrum without significant loss of accuracy.

\subsection{Downstream flow variables}
The spatial distribution of the flow variables, namely, pressure, density and velocity, are derived from the shock pressure evolution computed previously. For example, pressure perturbations downstream can be written in terms of Bessel functions as
\begin{equation}
\bar{p}\left(x,\tau\right)=\sum_{\nu=0}^{\infty}N_{\nu}J_{\nu}\left(\sqrt{\tau^2-x^2}\right)e^{-\nu\left[\tanh^{-1}\left(M_2\right)-\tanh^{-1}\left(\frac{x}{\tau}\right)\right]}\ .
\end{equation}

As the asymptotic expression \eqref{pstau} is found to reproduce accurately the shock pressure evolution, the asymptotic long-time response of the shock is employed to compute the post-shock disturbances. 

Downstream linear perturbations are conveniently split into entropic-vortical, conveyed by the fluid particles, and traveling acoustic modes \citep{Kovasznay53,Chu1958}, namely 
\begin{equation}
\begin{aligned}
&\bar{p}(x,\tau)= \bar{p}_a(x,\tau)\ ,&\quad &\bar{\rho}(x,\tau)= \bar{\rho}_a(x,\tau) + \bar{\rho}_e(x) \\
&\bar{u}(x,\tau)= \bar{u}_a(x,\tau)+\bar{u}_r(x)\ ,&\quad &\bar{v}(x,\tau)= \bar{v}_a(x,\tau) + \bar{v}_r(x)\ .\nonumber
\label{kova}
	\end{aligned}
\end{equation}

In the absence of diffusive effects, the amplitudes of the entropic-solenoidal perturbations are given by their corresponding values generated right behind the shock, and they are steady in a reference frame co-moving with the fluid particles. Acoustic disturbances, on the other hand, refer to traveling sonic waves that escape from the shock when $\zeta>1$.

The acoustic radiation condition is then determined by $\omega_s>(1-M_2^2)^{1/2}$, a condition that depends on the upstream shear wave, since $\zeta=\left[0,\infty\right)$ depends on the relative properties of the perturbation field ahead of the shock. Small values of $\zeta$ represent the interaction with upstream vortices highly stretched in the streamwise direction $\lambda_x\gg\lambda_y$, while the opposite is true for $\zeta\gg 1$. In the latter low mode-number scenario ($\lambda_x\ll\lambda_y$), the problem reduces to the one-dimensional interaction of the shock with radial perturbation waves. Such stability analysis has been developed by \citet{Velikovich2016} for the classical Noh's configuration in adiabatic conditions. The asymptotic far-field solution for the acoustic disturbances is also written in terms of harmonic functions, representing stable traveling fronts that occur only when the shock oscillation frequency is sufficiently high, $\zeta > 1$. 

Traveling sonic perturbations are functions of $(\omega_{a} \tau - k_{a} x)$, with the frequency $\omega_{a}$ and the wave number $k_{a}$ being determined by the post-shock adiabatic dispersion relationship $\omega_{a}^2=k_{a}^2+1$, and the shock oscillation frequency $\omega_{s}= \omega_{a} - M_2 k_{a}$, yielding
\begin{equation}
\omega_{a}=\frac{\omega_s - M_2\sqrt{\omega_s^2-1+M_2^2}}{1-M_2^2}
\label{omegaa}
\end{equation}
and
\begin{equation}
k_{a}=\frac{\omega_s M_2 - \sqrt{\omega_s^2-1+M_2^2}}{1-M_2^2}\ ,
\label{ka}
\end{equation}
respectively, which depend upon the shock frequency $\omega_s$. It is straightforward to see that $k_a$ can be either negative or positive, the former representing the sonic waves propagating downwards in the compressed gas reference frame, and the latter denoting the waves moving upwards, although never catching up the shock wave as dictated by the isolated-front boundary condition. The shock oscillation frequency, $\omega_s=1$, marks the standing acoustic wave regime, therefore separating the left-traveling solution $\omega_s>1$ from the right-traveling regime $(1-M_2^2)^{1/2}<\omega_s<1$ in the compressed gas reference frame. When the shock oscillates with two frequencies, the possibility of having sonic fronts running upstream and downstream is possible.

The asymptotic pressure and isentropic density perturbations, far behind the shock, equal 
\begin{equation}
\bar{p}(x,\tau)=\bar{\rho}_a(x,\tau)= \mathcal{P}_s \cos\left(\omega_a \tau-k_a x \right)\ ,
\label{pa}
\end{equation}
with $\mathcal{P}_s$ standing for the amplitude of the shock pressure disturbances in the long-wavelength regime. The amplitude of the associated acoustic-velocity perturbations are proportional to the pressure changes through the functions \eqref{UVa}, provided in the Appendix \ref{App1}. The corresponding isentropic temperature variations induced by the acoustic-shock radiation are simply $\bar{T}_a(x,\tau) = \left(\gamma-1\right) \bar{p}(x,\tau)$. 

The entropic contribution to the density perturbations $\bar{\rho}_e$ is computed from Rankine-Hugoniot relations~\eqref{massRH}-\eqref{eneRH}, after subtracting the acoustic part. It is readily seen that
\begin{equation}
\bar{\rho}_e(x)=\left(\mathcal{D}-1\right) \bar{p}_s\left(\tau=\frac{x}{M_2}\right)
\label{dene}
\end{equation}
with $\mathcal{D} = \left(2 M_2 \sigma_b -1\right)/M_2^2$ being the amplitude of the density perturbations behind the shock. As easily inferred from Fig.~\ref{fig:RH}, the value of $\mathcal{D}$ is found to be positive, and reaches a constant value in the strong-shock limit: $\mathcal{D}|_{M_1\gg1}=2 M_2^{-2}$. The corresponding isobaric temperature perturbation, scaled with base-flow temperature, is the function $\bar{T}_e(x) = -\bar{\rho}_e(x)= -(\mathcal{D}-1) \bar{p}_s(\tau=x/M_2)$.

Analogously, dimensionless vorticity disturbances are determined by
\begin{equation}
\bar{\omega}(x)=\frac{\partial \bar{v}}{\partial x}-\frac{\partial \bar{u}}{\partial y}=\Omega_2 \bar{p}_s\left(\tau=\frac{x}{M_2}\right) + \Omega_1  \cos\left(\frac{\omega_s}{M_2} x \right) 
\label{vort}
\end{equation}
with 
\begin{equation}
\Omega_1=C_2\left[1+\left(\frac{k_x}{k_y}\right)^2\right]=C_2\left(1+\frac{1-M_2^2}{C_2^2 M_2^2}\zeta^2\right)
\label{Omega1}
\end{equation}
indicating the contribution result of the one-dimensional compression effect, the shrinking of the vortices by the overall mass compression ratio, and 
\begin{equation}
\Omega_2=\frac{M_2 \left(C_2-1\right) \sigma_a+ \sigma_b M_2 -1}{M_2}
\label{Omega2}
\end{equation}
referring to contribution induced by shock rippling proportional to pressure, a two-dimensional effect.

The rotational contribution for the velocity disturbances is readily computed through the vorticity field, by knowing that rotational perturbations are steady and isobaric in the linear-inviscid approach. The relationships
\begin{equation}
\bar{\omega}(x)=-\frac{\partial^2 \bar{u}_r}{\partial x^2}+\bar{u}_r \ ,\ \bar{v}_r(x)=-\frac{\partial \bar{u}_r}{\partial x}
\label{urot}
\end{equation}
are then employed to write the asymptotic longitudinal and transverse rotational-velocity distributions, provided in eqs.~\eqref{urasym} and \eqref{vrasym}.

\begin{figure}
\includegraphics[width=0.47\textwidth]{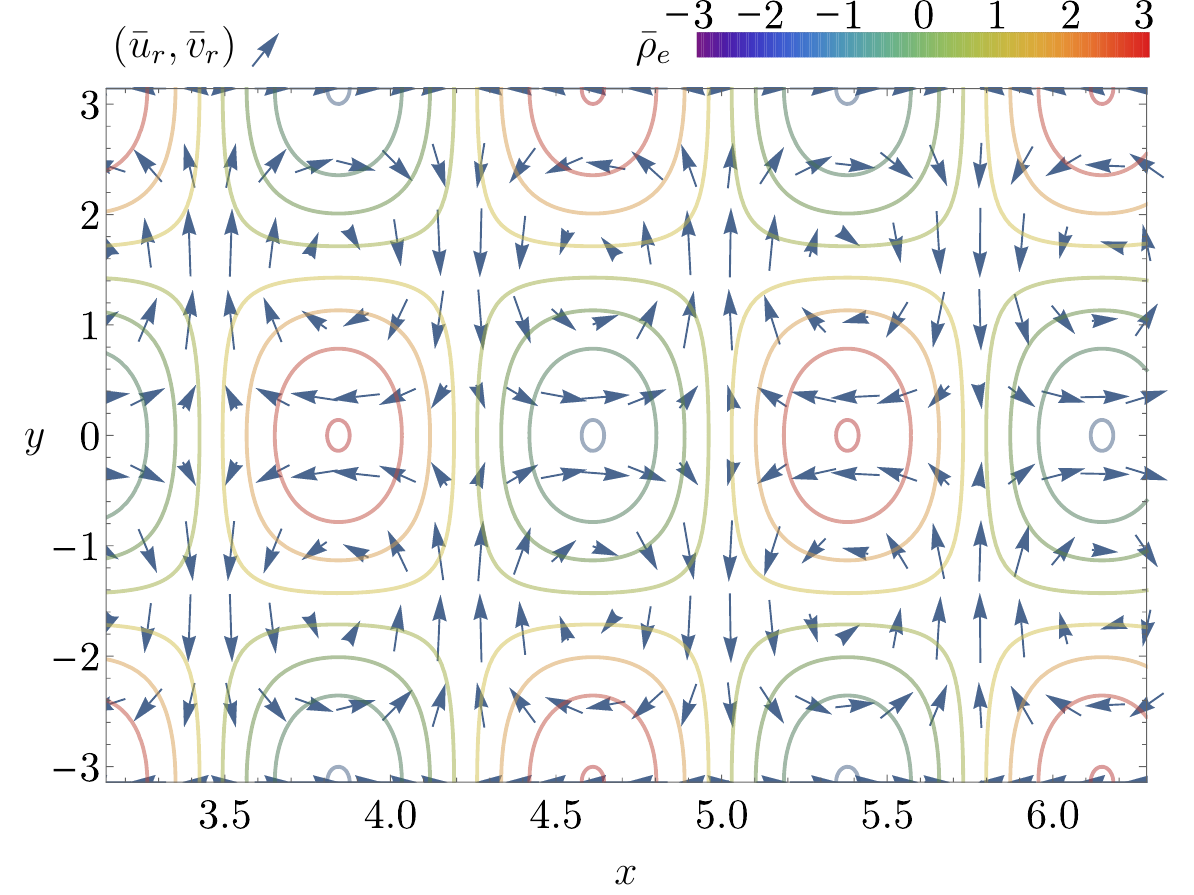}
\caption{Two-dimensional vector field plot for rotational-velocity perturbations superposed to iso-contours of entropic-density disturbances for a shock wave with $M_1=5$, $\varepsilon =0.4$, and $\zeta=1.2$.} 
\label{fig:voren}
\end{figure}

The asymptotic expressions for the rotational-velocity and entropic-density perturbations are computed in Fig.~\ref{fig:voren}, for the same conditions as in Fig. \ref{fig:pstau}. Velocity perturbations are displayed in a two-dimensional vector field, with the length of the vectors being scaled within the maximum and minimum velocity amplitudes, $1.9$ and $0.45$, respectively. Transverse component of the velocity perturbations is found to be much greater than longitudinal contribution. The spatial frequency modulation, given by $\omega_s/M_2$ is clearly distinguished. The amplitude of the rotational perturbations depends on the the incident angle $\theta$, as shown in Fig.~\ref{fig:UV} for $M_1=5$. This dependence is later used to account for the interaction with a whole spectrum of vorticity waves, with $\theta$  ranging from $0$ to $\pi$, upon consideration of the isotropic probability distribution.

Superposed to the vector field, the entropic-density disturbances are displayed in a contour plot in Fig.~\ref{fig:voren}. The center of the eddies and the peaks of the density field are shifted in $\pi/2$ in the lateral coordinate, as the former are proportional to $\sin(y)$ and the second to $\cos(y)$. Along the streamwise direction the peak values of density and rotational perturbations are in phase for $\zeta>1$, as both periodic distributions are proportional to $\cos{\left(\omega_s/M_2\, x\right)}$. There exits a spatial shift between the rotational and entropic mode, $\Delta \phi=\phi_r-\phi_e$, for $\zeta<1$, given by the contribution of the orthogonal components, $\tan \phi_r =\Omega_2\mathcal{P}_{li} /(\Omega_2\mathcal{P}_{lr}+\Omega_1)$ and $\tan \phi_e =\mathcal{P}_{li} /\mathcal{P}_{lr}$.

\section{Linear interaction analysis with 3D isotropic vorticity perturbations}
\label{sec:analysis_field}
\subsection{Turbulent kinetic energy}
\label{sec:tke}

The three-dimensional upstream flow is assumed to be homogeneous and isotropic. Therefore, the amplitude of the incident shear wave $\hat{u}_1$ depends exclusively on the wave-number amplitude $|\vec{k}|=k$ as $\vec{k}$ is uniformly distributed over the unit sphere. The three-dimensional problem is conveniently formulated in spherical polar coordinates, so the upstream velocity field $(\bar{u}_1,\bar{v}_1,\bar{w}_1)=\hat{u}_1 (\sin \theta \sin\varphi,\cos \theta \sin\varphi,\cos\varphi )$ and the associated wave-number vector is $\vec{k}=k (\cos \theta ,-\sin\theta,0 )$. The interaction with the whole spectrum of perturbations is carried out by direct superposition of linear perturbations \citep{Batchelor1953}. The average upstream velocity perturbation is
\begin{equation}
\langle\bar{u}_1^2\rangle= \int_{k^3}  |\bar{u}_1|^2{\rm d}k^3  = \frac{8\pi}{3} \int_0^{\infty}\hat{u}_1^2(k)k^2{\rm d}k\ ,
\label{uw3D}
\end{equation}
\begin{equation}
\langle\bar{v}_1^2\rangle= \langle\bar{w}_1^2\rangle= \int_{k^3}  |\bar{v}_1|^2{\rm d}k^3 =  \frac{2\pi}{3} \int_0^{\infty}\hat{u}_1^2(k)k^2{\rm d}k 
\label{vw3D}
\end{equation}
so the corresponding turbulent kinetic energy (TKE) computes as
\begin{equation}
\text{TKE}_1 = \frac{1}{2} \left( \langle\bar{u}_1^2 \rangle + \langle \bar{v}_1^2 \rangle + \langle \bar{w}_1^2 \rangle\right)= 2\pi\int_0^{\infty} \hat{u}_1^2(k) k^2 {\rm d}k
\label{TKEo}
\end{equation}
with $\hat{u}_1(k)=\text{fun}(k)$ representing the isotropic energy spectrum.

The problem is  further simplified by reducing the three-dimensional geometry into an equivalent two-dimensional case that accounts for the effect of vorticity perturbations that are parallel or perpendicular to the shock propagation velocity. After some straightforward algebra, the amplification ratio across the shock wave is 
\begin{equation}
K= \frac{\text{TKE}_2}{\text{TKE}_1} = \frac{1}{2}  \int_0^{\pi/2} \left(\bar{u}^2 + \bar{v}^2 \right)\sin^3\theta {\rm d}\theta + \frac{1}{2}
\label{K3Dtheta}
\end{equation}
which is conveniently rewritten in terms of the integration variable $\zeta$ as
\begin{equation}
\begin{aligned}
K=\frac{1}{3}  \int_0^{\infty} \left(\bar{u}^2 + \bar{v}^2 \right) \text{P}(\zeta) {\rm d}\zeta+ \frac{1}{2} 
\label{K3D}
\end{aligned}
\end{equation}
with
\begin{equation}
\text{P}(\zeta) = \frac{3}{2}\frac{M_2^4 C_2^4 \sqrt{1-M_2^2}}{\left[M_2^2 C_2^2 +\zeta^2 \left(1-M_2^2\right)\right]^{5/2}}
\label{pdf}
\end{equation}
standing for the normalized probability-density distribution obeying $\int_0^{\infty}\text{P}(\zeta){\rm d}\zeta=1$. It is readily seen that, although post-shock turbulence spectrum depends on upstream energy distribution $\int_0^{\infty}\hat{u}_1^2(k)k^2dk$, the kinetic energy amplification ratio does not as long as isotropic conditions are considered, namely $\hat{u}_1(k)=\text{fun}(k)$.

The amplification ratios for the longitudinal and transverse kinetic energy contributions can be computed with the aid of the probability density distribution. They are conveniently split into rotational and acoustic contributions, yielding
\begin{equation}
\begin{aligned}
L &= L_r + L_a =   \int_0^{1}\left[\left(\mathcal{U}_{li}^r\right)^2+\left(\mathcal{U}_{li}^r\right)^2\right] \text{P}(\zeta) {\rm d}\zeta+\\ 
&+\int_1^{\infty} \left(\mathcal{U}_{s}^r\right)^2  \text{P}(\zeta) {\rm d}\zeta + \int_1^{\infty} \left(\mathcal{U}^a\right)^2 \text{P}(\zeta) {\rm d}\zeta
\label{L3D}
\end{aligned}
\end{equation}
for the longitudinal part. The variation of the velocity perturbation amplitudes with $\zeta$ is deduced from Fig.~\ref{fig:UV} (for $M_1=5$) knowing that $\zeta$ is inversely proportional to $\tan\theta$, as \eqref{zeta} reads.

Equivalently, the turbulent kinetic energy associated to the transverse contribution is
\begin{equation}
\begin{aligned}
T &= T_r + T_a =  \frac{1}{2}  \int_0^{1}\left[\left(\mathcal{V}_{li}^r\right)^2+\left(\mathcal{V}_{li}^r\right)^2\right] \text{P}(\zeta) {\rm d}\zeta+\\ 
&+ \frac{1}{2}\int_1^{\infty} \left(\mathcal{V}_{s}^r\right)^2  \text{P}(\zeta) {\rm d}\zeta +  \frac{1}{2}\int_1^{\infty} \left(\mathcal{V}^a\right)^2 \text{P}(\zeta) {\rm d}\zeta+ \frac{3}{4}\ .
\end{aligned}
\label{T3D}
\end{equation}

The total turbulent kinetic energy, also split into 
rotational and acoustic contributions through $K=K_r+K_a$, is computed with the aid of $K_r= (L_r+2 T_r)/3$ and $K_a= (L_a+2 T_a)/3$, or equivalently through $K=\left(L+2T\right)/3$.

\begin{figure}
\includegraphics[width=0.47\textwidth]{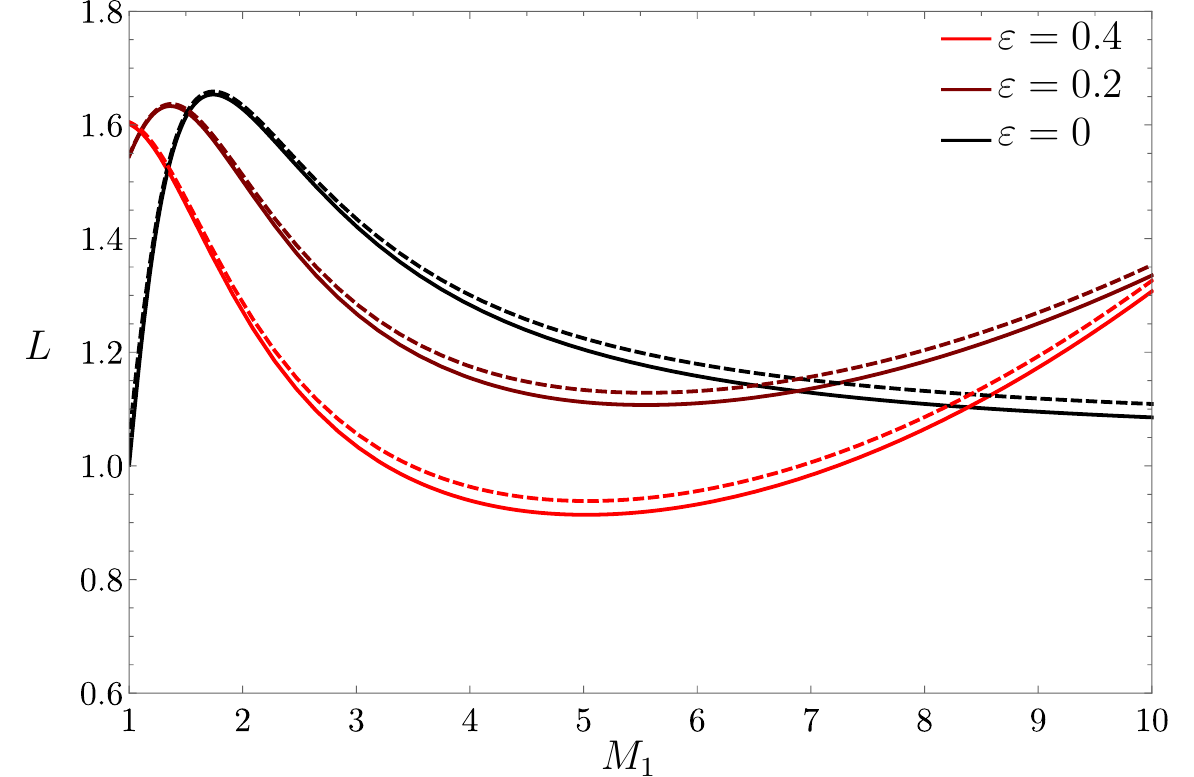} \vspace{2mm}
\\ \includegraphics[width=0.47\textwidth]{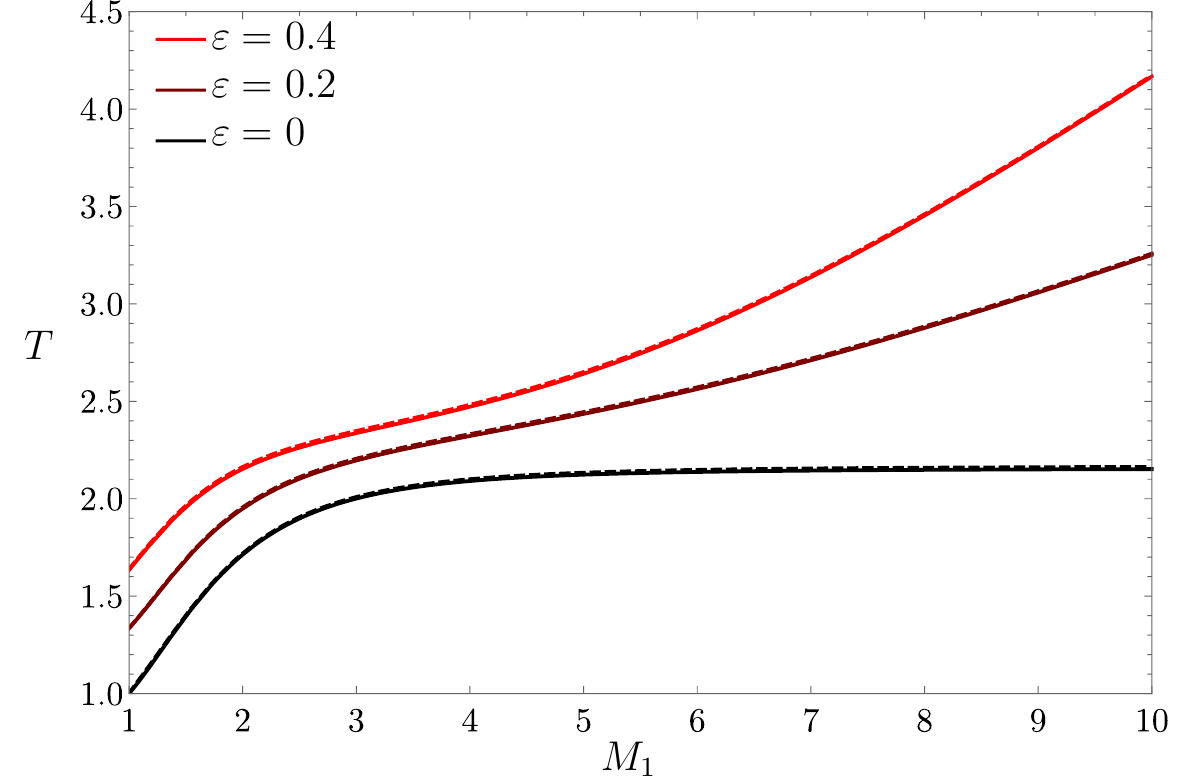}\vspace{2mm} \\ \includegraphics[width=0.47\textwidth]{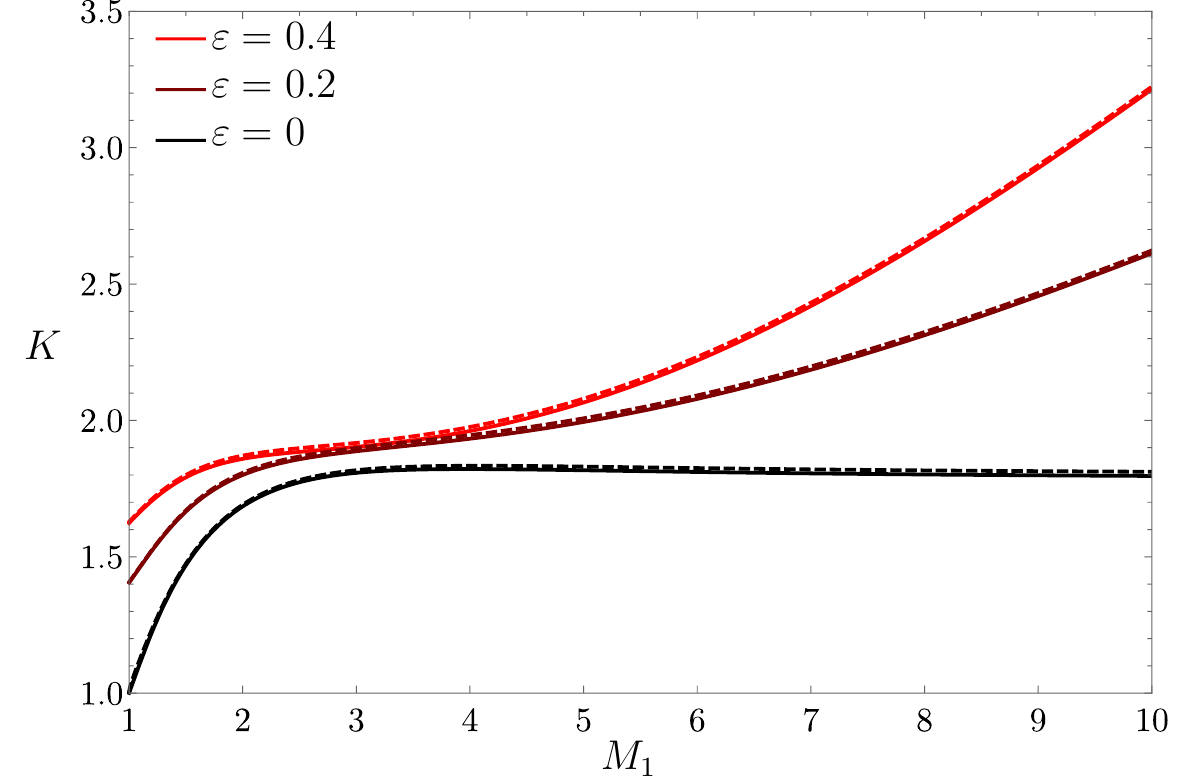}
\caption{Longitudinal $L$, transverse $T$ and total $K$ turbulent kinetic energy for $\varepsilon=$ 0, 0.2, and 0.4. The solid lines account for rotational contribution and the dashed lines show rotational and acoustic contributions.} 
\label{fig:LTK}
\end{figure}

The variation of the longitudinal, transverse and total contributions for the turbulent kinetic energy are shown in Fig.~\ref{fig:LTK} as a function of $M_1$, for $\varepsilon=$ 0, 0.2, and 0.4. The solid lines show the rotational contribution and the dashed lines include the contribution of both rotational and acoustic kinetic energy. In agreement with Fig.~\ref{fig:UV}, the acoustic contribution is found to be greater for the longitudinal part $L$, although sufficiently small to be neglected for any $M_1$ and $\varepsilon$ considered. Although not clearly seen in Fig.~\ref{fig:LTK}, the function $K$ approaches a constant value in the strong shock limit $M_1\gg1$, they are 1.8, 7.1, and 9.8 for $\varepsilon=$ 0, 0.2, and 0.4, respectively. On the other hand, the weak shock limit $M_1-1\ll1$ provides 1, 1.4, and 1.6 for the same conditions. For a fixed value of the incident Mach number, the effect of nuclear dissociation is seen to increase the total kinetic energy. It is found that, for a Mach number close to 3, the total kinetic energy is less sensitive to dissociation energy, although longitudinal and transverse contributions are clearly counter-affected. It indicates that post-shock anisotropy is modified by $\varepsilon$. Longitudinal contribution is generally diminished by nuclear dissociation if the Mach number is sufficiently high, a region that covers the scenarios of most interest. It is also found that transverse perturbations across the shock are more sensitive to the shock passage, then conforming a post-shock flow that differs from the ideal 1D configuration.

\begin{figure}
\includegraphics[width=0.47\textwidth]{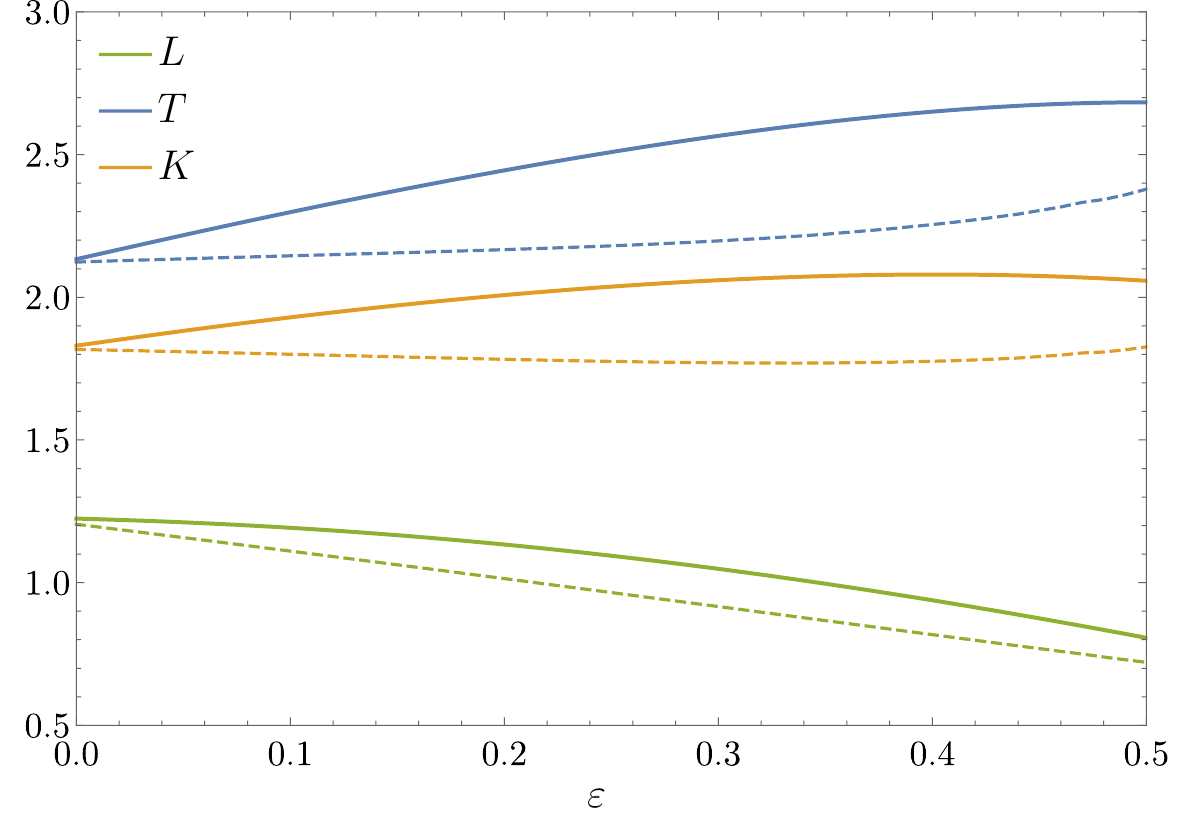}
\caption{Longitudinal $L$, transverse $T$ and total $K$ turbulent kinetic energy for $M_1=5$ as a function of $\varepsilon$. The solid lines represent computations of eqs.~\eqref{L3D}, \eqref{T3D} and \eqref{K3D}, while the dashed lines show the predictions in \protect\cite{Abdikamalov16}.} 
\label{fig:LTK1617}
\end{figure}

A direct comparison with the results obtained in \cite{Abdikamalov16} reveals that the dependence of the turbulent kinetic energy with $M_1$ and $\varepsilon$ is affected when endothermic effects are included in the linear perturbation analysis. Although similar trends, when increasing $\varepsilon$, is found in both works, the values may differ substantially when the energy employed in dissociating the gas is sufficiently high. For the sake of exemplification, predictions for $L$, $T$ and $K$ are computed in Fig. \ref{fig:LTK1617} by using eqs.~\eqref{L3D}, \eqref{T3D} and \eqref{K3D} (solid) and recasting the data in \cite{Abdikamalov16} (dashed). The differences become more pronounced with increasing shock strength, reaching $\sim 30\%$ in $K$ for $M_1 = 10$ and $\varepsilon=0.4$. 

\subsection{Turbulent Mach number}

In is instructive to relate the pre-shock and post-shock turbulent Mach numbers. It is immediate to see that
\begin{equation}
\langle \delta M_2^2\rangle = -4 M_2 \langle \bar{u}\bar{a} \rangle + \langle \bar{u}^2 \rangle + \langle \bar{v}^2 \rangle + \langle \bar{w}^2 \rangle +3 M_2^2 \langle \bar{a}^2\rangle\ ,
\end{equation}
which can be split into entropic-rotational and acoustic contributions as $\langle \delta M_2^2\rangle=\langle \delta M_1^2 \rangle \left( \Phi_{er}+\Phi_{ac}\right)$ in terms of functions $\Phi_{er}$ and $\Phi_{ac}$ represent these two contributions. For isotropic turbulence in the upstream flow, the entropic-rotational part reads
\begin{equation}
\begin{aligned}
\Phi_{er} &=\frac{M_2^2C_2^2}{M_1^2}\left[\frac{\langle \bar{u}_r^2 \rangle +  \langle \bar{v}_r^2 \rangle +  \langle \bar{w}_r^2 \rangle}{3\langle \bar{u}_1^2 \rangle}+ \frac{M_2^2}{4} \frac{\langle \bar{\rho}_e^2 \rangle}{\langle \bar{u}_1^2 \rangle} + \frac{2 M_2}{3}   \frac{\langle \bar{u}_r \bar{\rho}_e \rangle}{\langle \bar{u}_1^2 \rangle}\right]\\
 &= \frac{M_2^2C_2^2}{M_1^2}\left[K_r + \frac{M_2^2}{4} D_e+
\frac{2 M_2}{3} B_{er}\right]\ ,
\end{aligned}
\end{equation}
while the acoustic contribution can be expressed as
\begin{equation}
\begin{aligned}
\Phi_{ac} &=  \frac{M_2^2C_2^2}{M_1^2}\left[\frac{\langle \bar{u}_a^2 \rangle  +  \langle  \bar{v}_a^2 \rangle}{3\langle \bar{u}_1^2 \rangle}+ \frac{M_2^2}{4} \frac{\langle\bar{\rho}_a^2 \rangle}{\langle \bar{u}_1^2 \rangle}- \frac{2M_2}{3} \frac{\langle \bar{u}_a\bar{\rho}_a  \rangle}{\langle \bar{u}_1^2 \rangle} \right]\\
&=\frac{M_2^2C_2^2}{M_1^2}\left[K_a + \frac{M_2^2 (\gamma-1)^2}{4} D_a - \frac{2 M_2  (\gamma-1)}{3} B_a\right]\ .
\end{aligned}
\end{equation}
The values of $K_r$, $K_a$, $D_e$, $D_a$, $B_{er}$, and $B_a$ are provided in Eq.~\eqref{K3D} for the kinetic energy, in Eq.~\eqref{D3D} for the average density perturbations, and in Eq.~\eqref{B3D} for the buoyancy correlation. The mean value of the post-shock Mach number includes changes in the velocity field, density and the cross-product contribution. As $\bar{v}$ and $\bar{\rho}$ are orthogonal functions, only the longitudinal contribution correlates with density perturbations. The latter are being expressed as a function of shock pressure through $\bar{\rho}_e(x)= (\mathcal{D}-1)\bar{p}_s(\tau=x/M_2)$ for then entropic perturbations, and through $\bar{\rho}_a=\bar{p}_s(\tau=x/M_2)$ for the acoustic part.

\begin{figure}
\includegraphics[width=0.47\textwidth]{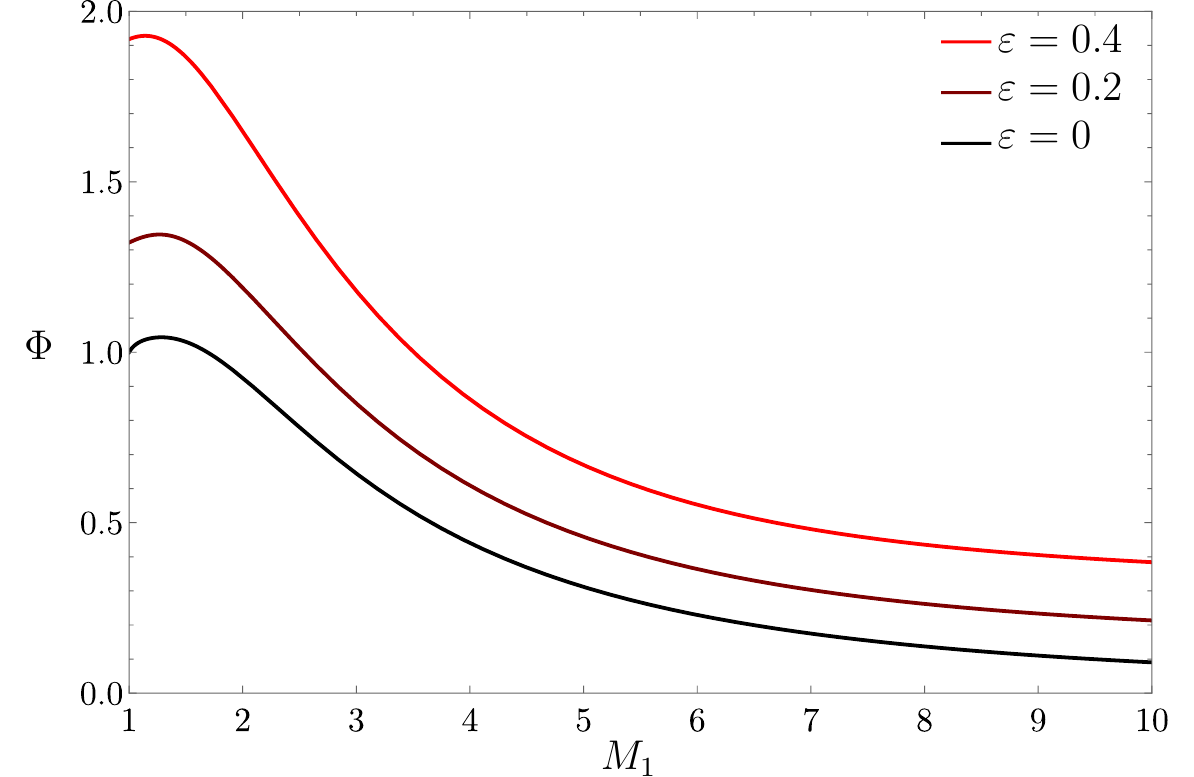}
\caption{Variable $\Phi$ as a function of the shock strength $M_1$ for $\varepsilon=$ 0, 0.2, and 0.4.} 
\label{fig:lum}
\end{figure}

The value of $\Phi=\Phi_{er}+\Phi_{ac}$ is computed in Fig.~\ref{fig:lum} as a function of the shock strength $M_1$ for $\varepsilon=$ 0, 0.2, and 0.4. For typical values these parameters ($0.2 \lesssim \varepsilon \lesssim 0.4$ and $M_1 \gtrsim 5$), $\Phi$ ranges from $\sim 0.3$ to $\sim 0.6$. Similarly to the turbulent kinetic energy in the post-shock region, most of the contribution to $\Phi$ comes from the entropic-rotational part, while the acoustic contribution $\Phi_{ac}$ is found to be negligibly small. 

\subsection{Enstrophy}
The effect of the shock passage on the upstream isotropic vorticity field can be computed with the aid of eqs. \eqref{vort} and \eqref{pdf}. The amplification of the average squared vorticity perturbations, nondimensionalized with $(k a_2)^2$, is written as
\begin{equation}
W = \frac{\langle\bar{\omega}_{x}^2+\bar{\omega}_{y}^2+\bar{\omega}_{z}^2\rangle}{\langle\bar{\omega}_{1,x}^2+\bar{\omega}_{1,y}^2+\bar{\omega}_{1,z}^2\rangle} = \frac{1}{3}+ \frac{2}{3}\frac{\langle \bar{\omega}_{y}^2+\bar{\omega}_{z}^2\rangle}{\langle\bar{\omega}_{1,y}^2+\bar{\omega}_{1,z}^2\rangle}=\frac{1}{3}+ \frac{2}{3}W_{\perp}\ ,
\label{W3D}
\end{equation}
with the factor $1/3$ referring to the invariable component of the vorticity pointing in the streamwise direction, and $W_{\perp}$ being the amplification factor of the averaged squared vorticity perpendicular to the shock propagation velocity. The two-dimensional equivalent factor
\begin{equation}
\begin{aligned}
&W_z = \frac{\langle \bar{\omega}_{z}^2\rangle}{\langle \bar{\omega}_{1,z}^2\rangle}=\int_1^\infty\left(\Omega_1+\Omega_2\mathcal{P}_{s}\right)^2\frac{C_2^2 M_2^2}{C_2^2 M_2^2+(1-M_2^2)\zeta^2} \text{P}(\zeta) {\rm d}\zeta \\
&+ \int_0^1\left[\left(\Omega_1+\Omega_2\mathcal{P}_{lr}\right)^2+\Omega_2^2\mathcal{P}_{li}^2\right]\frac{C_2^2 M_2^2}{C_2^2 M_2^2+(1-M_2^2)\zeta^2} \text{P}(\zeta) {\rm d}\zeta 
\label{Wz}
\end{aligned}
\end{equation}
is conveniently employed in computing the perpendicular contribution as $W_{\perp}=(C_2+3W_z)/4$. 

\begin{figure}
\includegraphics[width=0.47\textwidth]{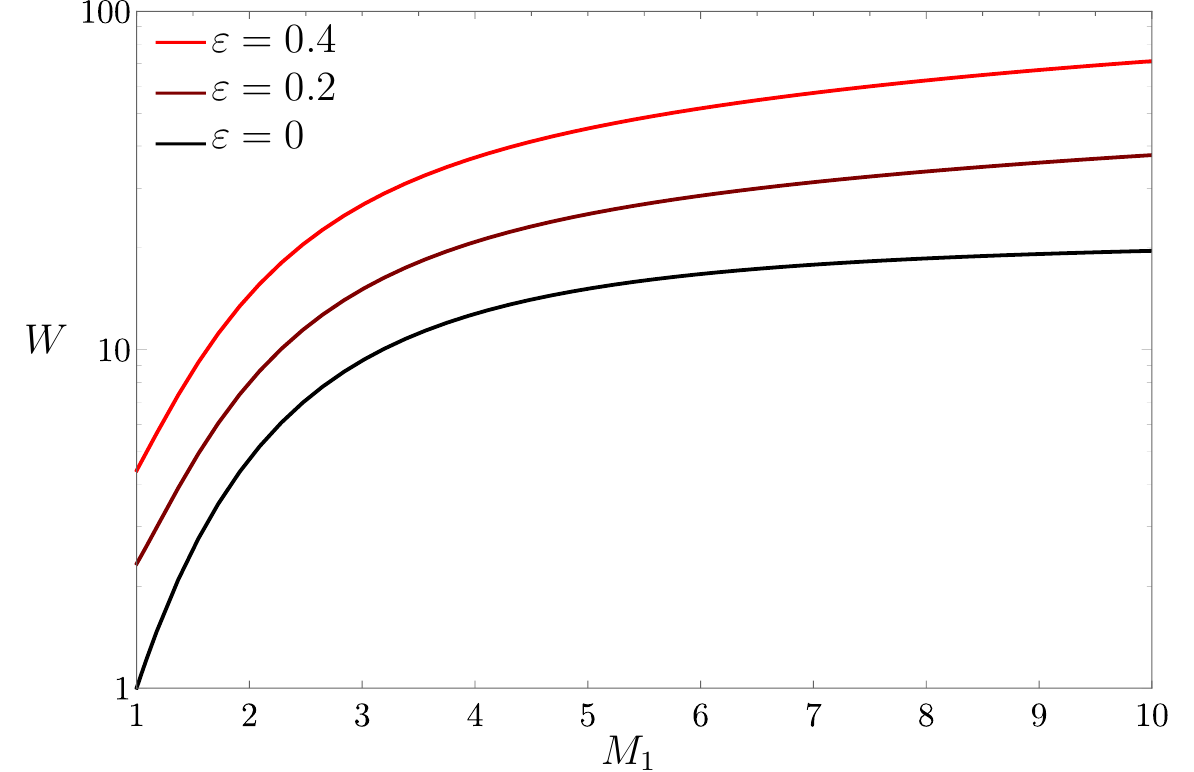}
\caption{Mean value of squared vorticity perturbations, $W$, for $\varepsilon=$ 0, 0.2, and 0.4.} 
\label{fig:W}
\end{figure}

The so-called enstrophy, $W$, is computed in  Fig.~\ref{fig:W} for the same conditions as in Fig.~\ref{fig:LTK}. In consonance to the turbulent kinetic energy, the effect of nuclear dissociation across the shock is found to increase the average vorticity intensity, for a fixed value of $M_1$.

When the shock is expanding at variable Mach number, the theory still holds if base-flow changes are negligible within the perturbation wavelength distance. Upstream turbulent flows characterized by short wavelengths will meet this constriction. On the other side, perturbations must be sufficiently large for the shock to be seen as a pure discontinuity. In such case, the post-shock kinetic energy at any radial locus can be approximated by the one left by the expanding shock, whose instantaneous properties $M_1$ and $\varepsilon$ can be computed following the analysis presented in next section. The values obtained for the downstream kinetic energy and enstrophy can be then used to compute the evolution of the turbulent flow by viscous-dissipative effects. Then, Figs.~\ref{fig:LTK} and \ref{fig:W} serve as the onset for such post-shock stage, with the subsequent thermalization of the kinetic energy being inferred by the dissipation energy cascade associated to the dominant scales \citep{Mabanta17}.

\section{Nuclear Dissociation Energy and the Pre-shock Mach Number In CCSN models} \label{sec:varepsilon}

\begin{figure}
\includegraphics[width=0.47\textwidth]{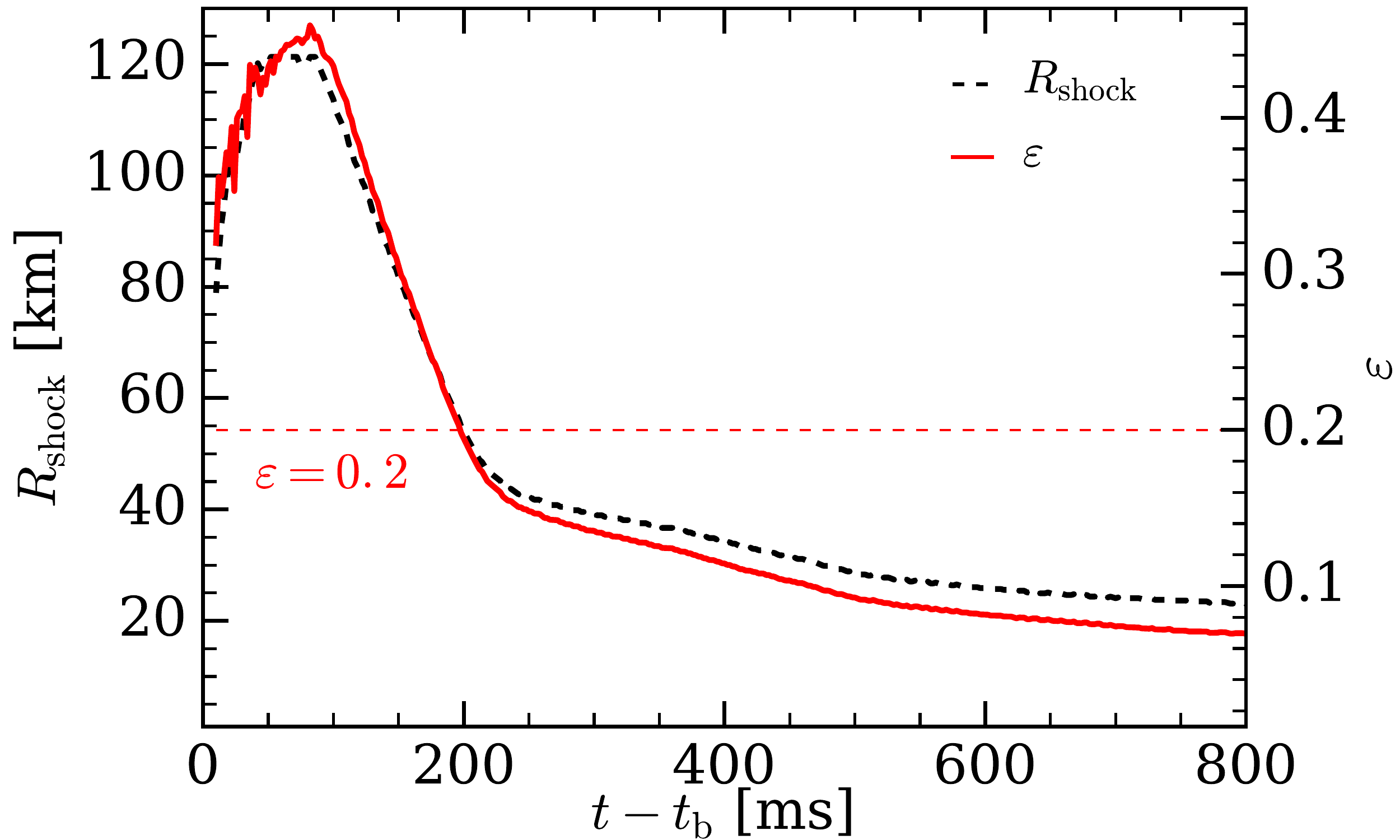}
\includegraphics[width=0.47\textwidth]{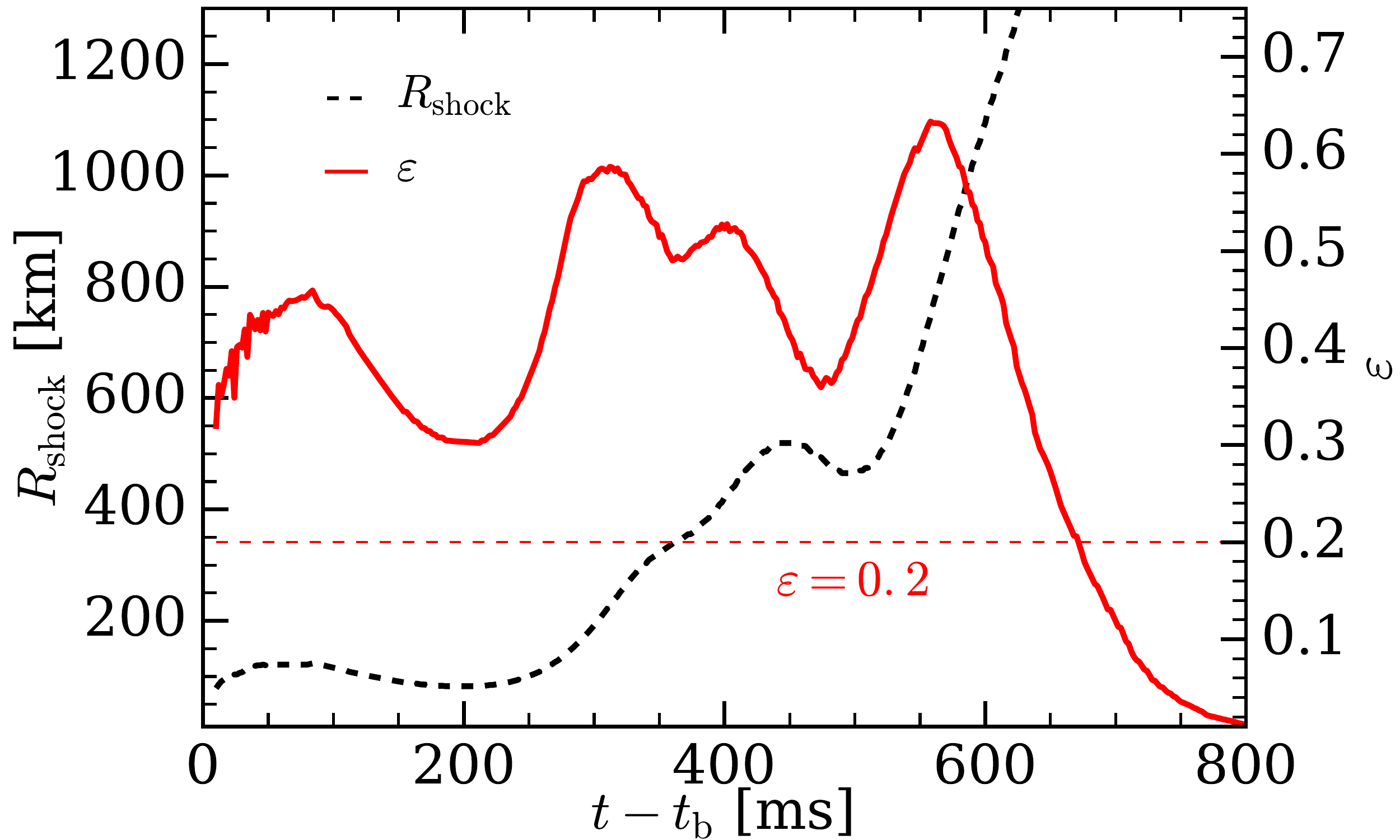}
\includegraphics[width=0.47\textwidth]{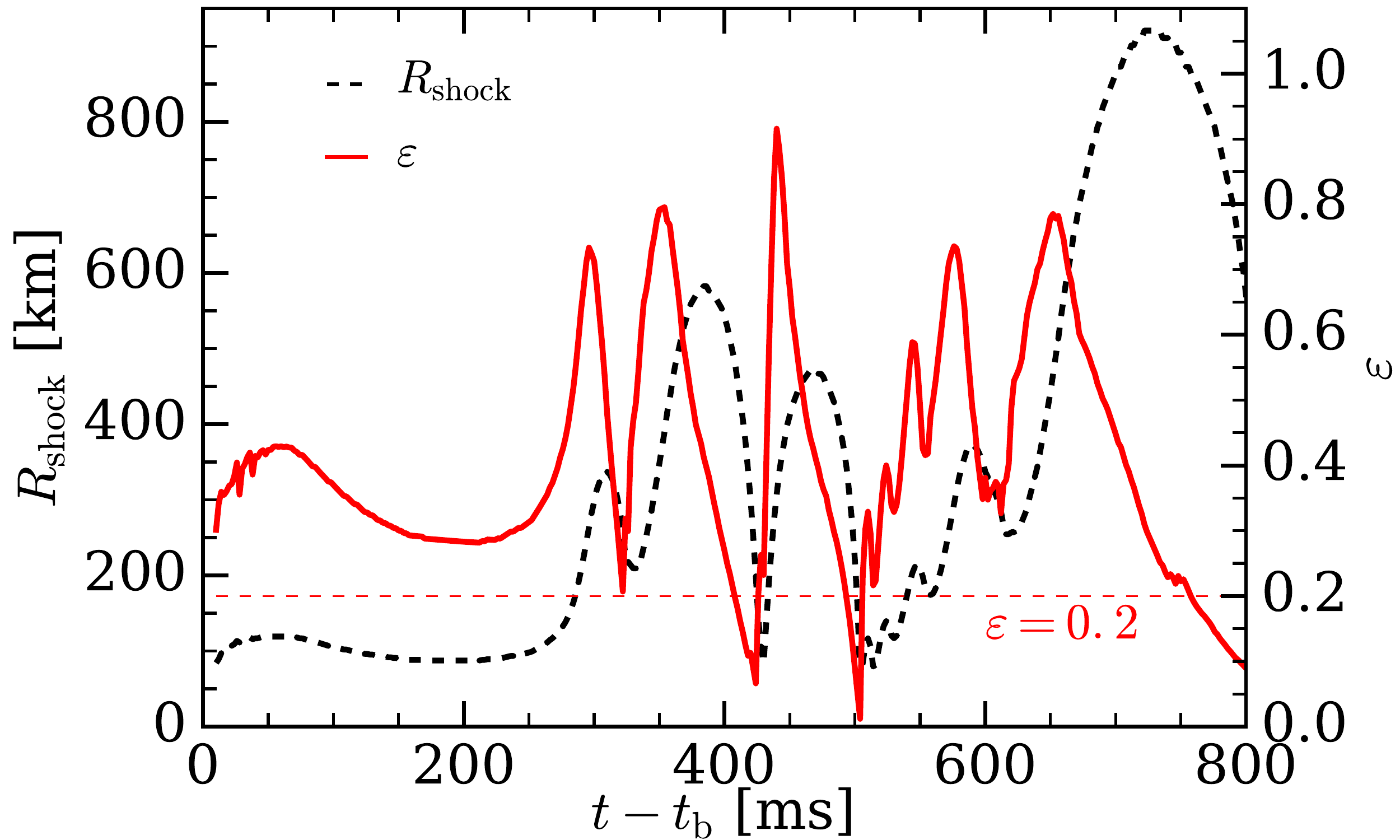}
\caption{{\bf Top panel:} Time evolution of the shock radius (dashed black line) and the nuclear dissociation parameter (solid red line) for non-exploding model $s15$ with heating factor $h=1.2$ (i.e., a group I model). For reference, the horizontal red dashed line shows the $\varepsilon=0.2$ line. {\bf Center panel:} The same as in top panel but for exploding model $s15$ with heating factor $h=1.23$ (i.e., a group II model). {\bf Bottom panel:} The same as in top panel but for model $s25$ with heating factor $h=1.18$ that undergoes strong shock oscillations (i.e., a group III model).} 
\label{fig:rsh_e_m}
\end{figure}

\begin{figure}
\includegraphics[width=0.47\textwidth]{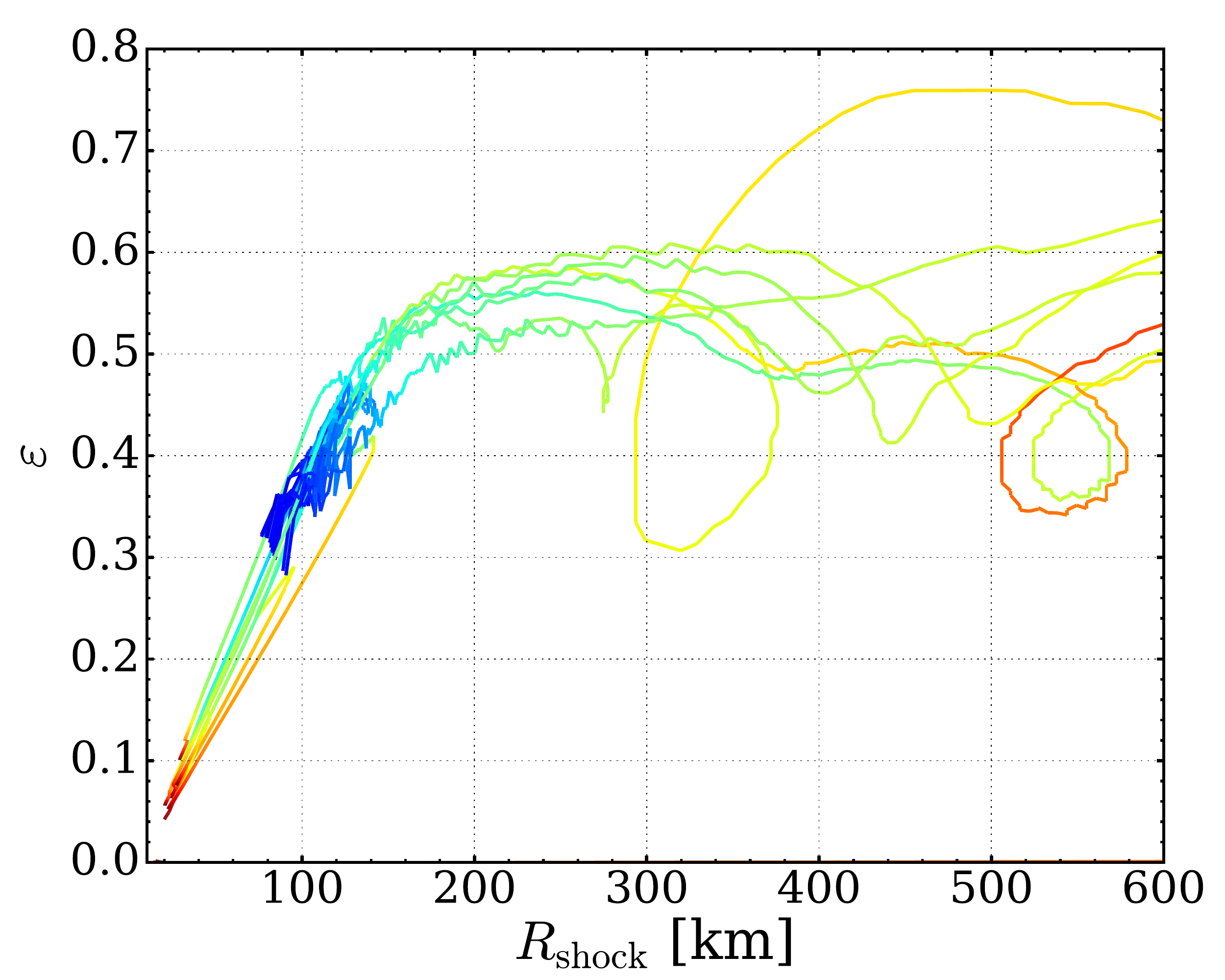}
\caption{The nuclear dissociation parameter $\varepsilon$ as a function of the shock radius for non-exploding (group I) and exploding (group II) models. Each line represents a specific model and the color of each line indicates the time: the blue end of the line corresponds to $10\,\mathrm{ms}$ after bounce, while the red end corresponds to late postbounce time ($t-t_\mathrm{b} \sim 1\,\mathrm{s}$). For shock radii $R_\mathrm{shock}\lesssim 175 \,\mathrm{km}$, $\varepsilon$ scales as $\propto R_\mathrm{shock}$, while for large shock radii, the growth of $\varepsilon$ saturates and remains $\sim 0.5$ until $R_\mathrm{shock}\lesssim 600 \,\mathrm{km}$.}
\label{fig:e_vs_rsh2}
\end{figure}

This section presents the estimates of the nuclear dissociation energy and the pre-shock Mach number from a series of spherically-symmetric CCSN simulations using the {\tt GR1D} code with the leakage/heating scheme \citep{Oconnor10}. Eight \cite{Woosley07} progenitor star models with ZAMS masses of $12M_\odot$, $15M_\odot$, $18M_\odot$, $20M_\odot$, $25M_\odot$, $30M_\odot$, $40M_\odot$, and $70M_\odot$ were considered. Each progenitor model is evolved using several values of the heating parameter. This yields a variety of qualitatively different evolutionary paths for each stellar model, ranging from non-exploding models to rapidly exploding models.  Each simulation is named using the following convention: for example, the simulation $s15h1.23$ uses a progenitor model with a ZAMS mass of $15M_\odot$ evolved with heating factor of $1.23$ \citep[for the definition of the heating factor, see, e.g.,][]{Oconnor10,Ott13}. 

Our simulations use the SFHo finite-temperature nuclear EOS of \cite{Steiner13}\footnote{Available at {\tt www.stellarcollapse.org} \citep{Oconnor10}.} as this EOS employs an accurate treatment of light nuclei. Calculations with the \cite{Lattimer91} EOS with nuclear incompressibility of $K=220$ MeV revealed similar results. Across our computational domain, we use $1000$ logarithmic radial grids with the central resolution of $0.1\,\mathrm{km}$. The outer boundary is fixed at the radius where the initial density is $2\times10^3 \, \mathrm{g/cm^3}$. 

The shock wave dissociates heavy nuclei into light nuclei such as $\alpha$ particles and free nucleons. The SFHo EOS includes the nuclei ${}^2$H , ${}^3$H, ${}^3$H, ${}^3$He, ${}^4$Li, $\alpha$ particles, and heavy nuclei. Based on the change of the mass fractions of nuclei across the shock, the nuclear dissociation parameter is calculated using formula (\ref{eq:varepsilon2}) derived in Appendix~\ref{App2}. The binding energies of the light nuclei are taken from the \cite{Audi03} database, while that of heavy nuclei are assumed to be equal to that of iron nuclei, i.e., $8.8$ MeV per nucleon. For calculating the dissociation energy at the shock, this is a reasonable assumption as the binding energies of heavy nuclei in the iron core and Si/O shells differ by at most $\sim 10\%$. 

The qualitative behaviors of $\varepsilon$ and $M_1$ depends on the overall dynamics of each model. In this respect, all the models considered here can be categorized into three groups: ($i$) non-exploding models, in which the shock wave gradually decreases with time without exhibiting strong radial oscillations (group I), ($ii$) exploding models, in which the shock gradually expands without strong oscillations (group II), and ($iii$) models, in which the shock wave exhibits strong oscillations before either transitioning to explosion or failing to explode (group III). In the following, we describe these three different model groups separately. 

The top panel of Fig.~\ref{fig:rsh_e_m} shows the shock radius (solid black line) and the dissociation parameter $\varepsilon$ (solid red line) as a function of post-bounce time for model $s15h1.22$. This is a non-exploding model, in which the shock gradually recedes without exhibiting strong radial oscillations, i.e., this model belongs to group I. After the initial period of $\sim 50\,\mathrm{ms}$, during which shock undergoes rapid expansion, the shock stalls until $t-t_\mathrm{b}\sim 100\,\mathrm{ms}$, after which $R_\mathrm{shock}$ starts receding monotonically. The qualitative behavior of $\varepsilon$ is similar to that of $R_\mathrm{shock}$: following the initial period of increase and subsequent stagnation, $\varepsilon$ gradually decreases with time. The dissociation parameters $\varepsilon$ falls below, e.g., $\varepsilon=0.2$ when $R_\mathrm{shock} \lesssim 55\,\mathrm{km}$. Other models of group I exhibit a similar behavior. 

The center panel of Fig.~\ref{fig:rsh_e_m} shows the shock radius (dashed black line) and the dissociation parameter $\varepsilon$ (solid red line) as a function of post-bounce time for model $s15h1.23$. This is an exploding model, in which the shock gradually expands without exhibiting strong radial oscillations, i.e., it belongs to group II. In this model, the stalled shock phase lasts until $t-t_\mathrm{b}\sim 200\,\mathrm{ms}$, after which $R_\mathrm{shock}$ slowly increases. In this phase, $R_\mathrm{shock}$ exhibits only weak oscillations with a relative amplitude of a few percent. At $t-t_\mathrm{b}\sim 500\,\mathrm{ms}$, the shock starts rapidly expanding and the model quickly transitions towards explosion. In the early $t-t_\mathrm{b}\lesssim 500\,\mathrm{ms}$ after bounce, the dissociation parameter stays above $0.2$ and oscillates around the value of $\sim 0.5$. However, it rapidly decreases during the explosion phase, once the shock radius becomes $\gtrsim 800\,\mathrm{km}$.  Other models of group II exhibit a similar behavior. 

It is illuminating to analyze $\varepsilon$ as a function of shock radius, a plot of which is shown in Fig.~\ref{fig:e_vs_rsh2} for all of our models in group I and II. Each line in this plot corresponds to one model and the color of a point on this line reflect that the time after bounce: the blue end of each line corresponds to $t-t_\mathrm{b}=10\,\mathrm{ms}$, while the red part corresponds to the end of the simulations ($t-t_\mathrm{b}\sim 1 \,\mathrm{s}$). In all non-exploding models (group I), $\varepsilon$ scales as $\propto R_\mathrm{shock}$, with the proportionality depending on mass:
\begin{equation}
\varepsilon \sim \frac{2}{3} M_{1.3}^{-1} \left( \frac{R_\mathrm{shock}}{150\,\mathrm{km}} \right),
\label{eq:eps_scaling}
\end{equation}
This relation is qualitatively similar to Eq.~(4) predicted by \citet{Fernandez2009a}. However, as can be seen in Fig.~\ref{fig:e_vs_rsh2}, the $\varepsilon\propto R_\mathrm{shock}$ scaling becomes invalid as soon as shock becomes larger than $\sim 175\,\mathrm{km}$, which occurs in exploding models. In this regime, $\varepsilon$ stops growing with $R_\mathrm{shock}$ and saturates to $\sim 0.5$ for most models. 

\begin{figure}
\includegraphics[width=0.47\textwidth]{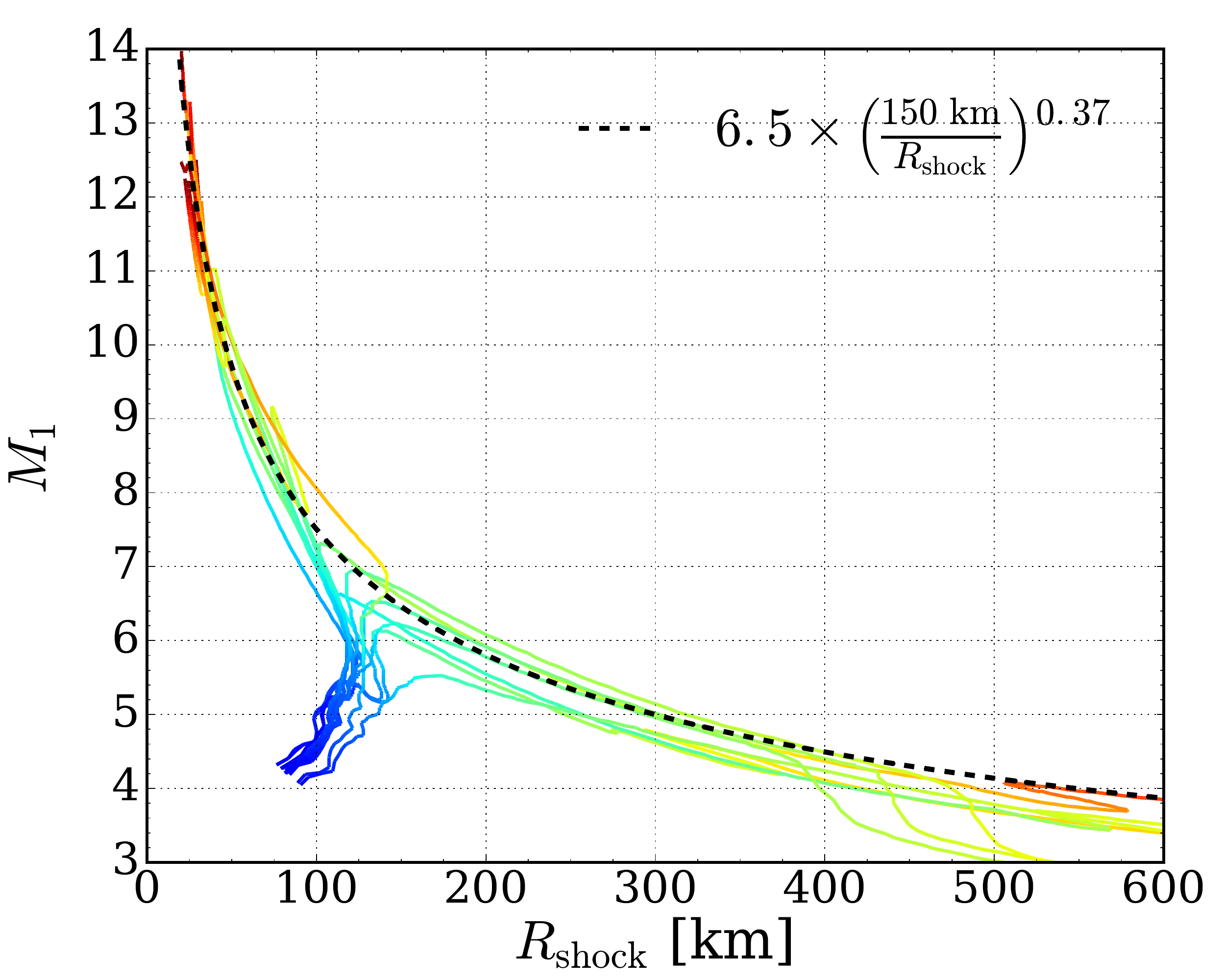}
\caption{Mach number as a function of the shock radius. The color of each line indicates the corresponding post-bounce time at which this value of the Mach number is extracted. The blue end of the lines corresponds to early post-bounce time of $t-t_\mathrm{b} = 10 \, \mathrm{ms}$, while the red region corresponds to late post-bounce time ($t-t_\mathrm{b} \sim 1 \, \mathrm{s}$). The dashed black line represents fitting function (\ref{eq:M1_scaling}) that yield the values of the pre-shock Mach number as a function of the shock radius $R_\mathrm{shock}$ in the stalled shock phase.}
\label{fig:m_vs_rsh}
\end{figure}

Figure~\ref{fig:m_vs_rsh} shows pre-shock Mach number $M_1$ as a function of shock radius for all of our models in groups I and II. As in Fig.~\ref{fig:e_vs_rsh2}, each line represents a single model and the color of each point on each line represents the post-bounce time. Except the immediate post-bounce time ($t-t_\mathrm{b}\sim 10-20\,\mathrm{ms}$), $M_1$ depends on $R_\mathrm{shock}$ as 
\begin{equation}
M_1 \sim 6.5\times \left( \frac{150\,\mathrm{km}}{R_\mathrm{shock}} \right)^{0.37}
\label{eq:M1_scaling}
\end{equation}
This relations is only approximate and the spread of the values of $M_1$ at a given $R_\mathrm{shock}$ is caused by the fact that different models have somewhat different thermodynamic conditions (e.g., temperature), which leads to different values of the speed of sound, which, in turn, affects the Mach number. 

Finally, the bottom panel of Fig.~\ref{fig:rsh_e_m} shows the shock radius (solid black line) and the nuclear dissociation parameter $\varepsilon$ (solid red line) as a function of time for models $s25h1.18$. This model exhibits strong radial shock oscillations from $\sim200\,\mathrm{ms}$ till $\sim800\,\mathrm{ms}$ after bounce. During this time, $\varepsilon$ also undergoes strong oscillations with the same frequency as the shock radius. The oscillations in the two quantities are somewhat out of phase. When the increase of $R_\mathrm{shock}$ is decelerating, $\varepsilon$ starts decreasing fast, reaching its local minimum just before $R_\mathrm{shock}$ does. It starts increasing when the shock radius is approaching its local minimum. At its minimum, $\varepsilon$ can become as small as $0.1$ for a brief period of time. The frequency of these oscillations are comparable to the frequencies of the infalling perturbations. For this reason, the linear formalism presented in this work is unlikely to be applicable to such models (cf. Section~\ref{sec:perturb_problem}). On the other hand, such oscillations are artificially strong in 1D models. Full 3D simulations are unlikely to exhibit strong oscillations, at least not in the angle-averaged shock radius. However, in the presence of strong SASI oscillations, the shock radius may oscillate along radial directions. In these situations, the values of $\varepsilon$ are likely exhibit similar oscillations as in models in group III.  

\subsection{Amplification of turbulent kinetic energy as a function of the shock radius}

\begin{figure}
\includegraphics[width=0.45\textwidth]{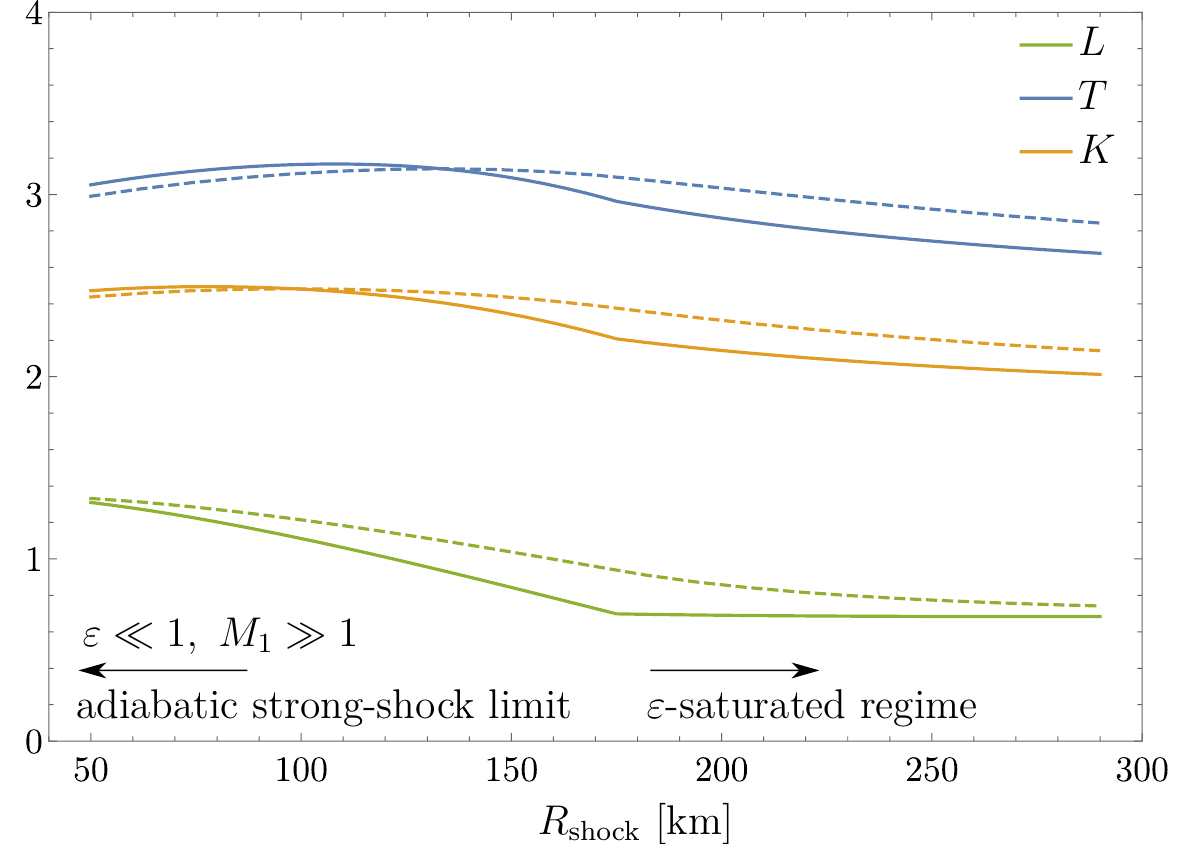}
\caption{Longitudinal $L$, transverse $T$ and total $K$ turbulent kinetic energy for $M_1=5$ as a function of the shock radius. The dashed lines represent the amplification of the integration kinetic energy in the post-shock region (cf. Eq.~\ref{eq:LTK_int}).} 
\label{fig:LTKrs}
\end{figure}

In addition to analyzing the amplification of turbulent kinetic energy across the shock as a function of parameters $\varepsilon$ and $M_1$, as was done in Section~\ref{sec:analysis_field}, one can get additional insight by looking at it as a function of the shock radius $R_\mathrm{shock}$. To this end, equations \eqref{eq:eps_scaling} and \eqref{eq:M1_scaling} allows us to express the nuclear dissociation degree $\varepsilon$ and shock strength $M_1$ as functions of the shock radius, $R_\mathrm{shock}$. These expressions are employed to compute $L$, $T$ and $K$ as a function of $R_\mathrm{shock}$ in Fig.~\ref{fig:LTKrs}. Each component of the turbulent kinetic energy appears to depend rather weakly on $R_\mathrm{shock}$. The transverse component increases by a factor of $\sim 3$, while the longitudinal component experiences no significant amplification. The total turbulent kinetic energy amplifies by a factor of $\sim 2$. As dictated by computations in Fig.~\ref{fig:e_vs_rsh2}, there exit two distinguished regions, the zone where $\varepsilon$ is linearly proportional to the shock position ($R_s\leq$175 km) and the region where nuclear dissociation is saturated. For small radius, the strong-shock adiabatic limit applies, as $M_1$ grows proportional to $R_s^{-0.37}$ and $\varepsilon$ approaches to zero. 

The dashed lines in Fig.~\ref{fig:LTKrs} represent the amplification of the integrated kinetic energy in the region of space confined between the shock and the center through 
\begin{equation}
\label{eq:LTK_int}
\begin{pmatrix} \bar L\\ \bar T\\ \bar K \end{pmatrix}
 = \frac{3}{R_{\text{shock}}^3}\int_0^{R_{\text{shock}}}\begin{pmatrix} L(r)\\ T(r)\\ K(r) \end{pmatrix}r^2 \rm{d} r\ ,
\end{equation}
provided that the characteristic time of post-shock turbulent structures evolution due to viscous-diffusive effects is much longer than shock-time passage through the matter to the distance $R_{\text{shock}}$.

\section{Discussion: Impact on the CCSN Explosion Mechanism}
\label{sec:discussion}

Generally speaking, the pre-shock perturbations in CCSNe consist of different physical modes, including acoustic and entropy waves in addition to the vorticity modes considered in this work. Without including all of these modes, one cannot obtain a rigorous estimate of the impact of perturbations on the explosion condition. However, in the linear order and for uniform mean flow, all these modes evolve independently from each other. Therefore, we can study the effect of vorticity modes alone in this work. The effect of other modes will be given in a future work. 

The impact of the perturbations on the explosion condition can be analyzed using the concept of the critical neutrino luminosity, i.e., the minimum neutrino luminosity that is necessary in order to produce an explosion for a given stellar model \citep{Burrows93}. The turbulence behind the supernova shock reduces the critical luminosity, an analytical estimate of which was obtained by \citet{Mueller15}:
\begin{equation}
\label{eq:crit_lum}
L_\mathrm{crit} \propto \left(\dot M M\right)^{3/5}r_\mathrm{gain}^{-2/5}\left(1+\frac{4}{3}\langle \delta M_2^2\rangle \right)^{-3/5}, 
\end{equation}
where $\delta M_2$ is the turbulent Mach number in the gain region. It is comprised of two contributions, one coming from neutrino-driven convection and/or SASI, another stemming from the perturbations crossing the shock. \citet{Mueller15} argue that the impact of the density perturbations generated by the advection of vorticity waves plays the dominant role in driving buoyancy-driven turbulence in the post-shock region. The resulting reduction in the critical luminosity was recently calculated by \citet{Mueller16}: 
\begin{equation}
\begin{aligned}
\label{eq:dl}
\frac{\Delta L_\mathrm{crit}}{L_\mathrm{crit}} \simeq - \frac{0.15 \pi}{l \eta_\mathrm{acc}\eta_\mathrm{heat}} \sqrt{\langle \delta M^2_0 \rangle},
\end{aligned}
\end{equation}
where $\delta M_0$ is the turbulent Mach number in the convective nuclear burning shell prior to collapse, $l$ is the angular wavenumber of the dominant perturbation, $\eta_\mathrm{heat}$ and $\eta_\mathrm{acc}$ are the efficiencies of neutrino heating and accretion. 

Expression (\ref{eq:dl}) is derived under the assumption that the advection of convective perturbations from Si/O shells towards the shock generates density perturbations of order
\begin{equation}
\label{eq:drho2_m2}
\sqrt{\langle \bar\rho^2_2 \rangle} \sim \sqrt{\langle \delta M_0^2 \rangle}
\end{equation}
behind the shock. This estimate does not include the density fluctuations associated with entropy perturbations generated in the post-shock region by the interaction of the shock with vorticity waves. Below, we estimate the impact of these perturbations on the critical luminosity. 

\subsection{Density perturbations in the post-shock region}
\label{sec:density_pert}

According to the linearized RH equations, \eqref{massRH}-\eqref{eneRH}, the corrugated shock front induces density perturbations in the post-shock gas. Such perturbations are of entropic ($\hat{\rho}_e$) and acoustic ($\hat{\rho}_a$) nature, with the former remaining frozen to the fluid particles in the absence of diffusive effects. For isotropic field of incoming vorticity perturbations, the average of the squared density changes in the post-shock region can be written as
\begin{equation}
\begin{aligned}
\langle\bar{\rho}^2\rangle = D \int_{0}^\infty\hat{u}_1^2(k)k^2dk
\label{denave}
\end{aligned}
\end{equation}
with the dimensional pre-spectrum coefficient $D$, split into entropic $D_e$ and acoustic $D_a$ contributions, being computed as
\begin{equation}
\begin{aligned}
D &= D_e + D_a =   \left(\mathcal{D}-1\right)^2\int_0^{1}\left(\mathcal{P}_{li}^2+\mathcal{P}_{li}^2\right) \text{P}(\zeta) {\rm d}\zeta+\\ 
&+\left(\mathcal{D}-1\right)^2\int_1^{\infty}\mathcal{P}_{s}^2 \, \text{P}(\zeta) {\rm d}\zeta + \int_1^{\infty} \mathcal{P}^2\, \text{P}(\zeta) {\rm d}\zeta\ .
\label{D3D}
\end{aligned}
\end{equation}
The terms involving the factor $\left(\mathcal{D}-1\right)^2$ correspond to the entropic contribution $D_e$, while the last term refers to the acoustic part $D_a$. 

\begin{figure}
\includegraphics[width=0.47\textwidth]{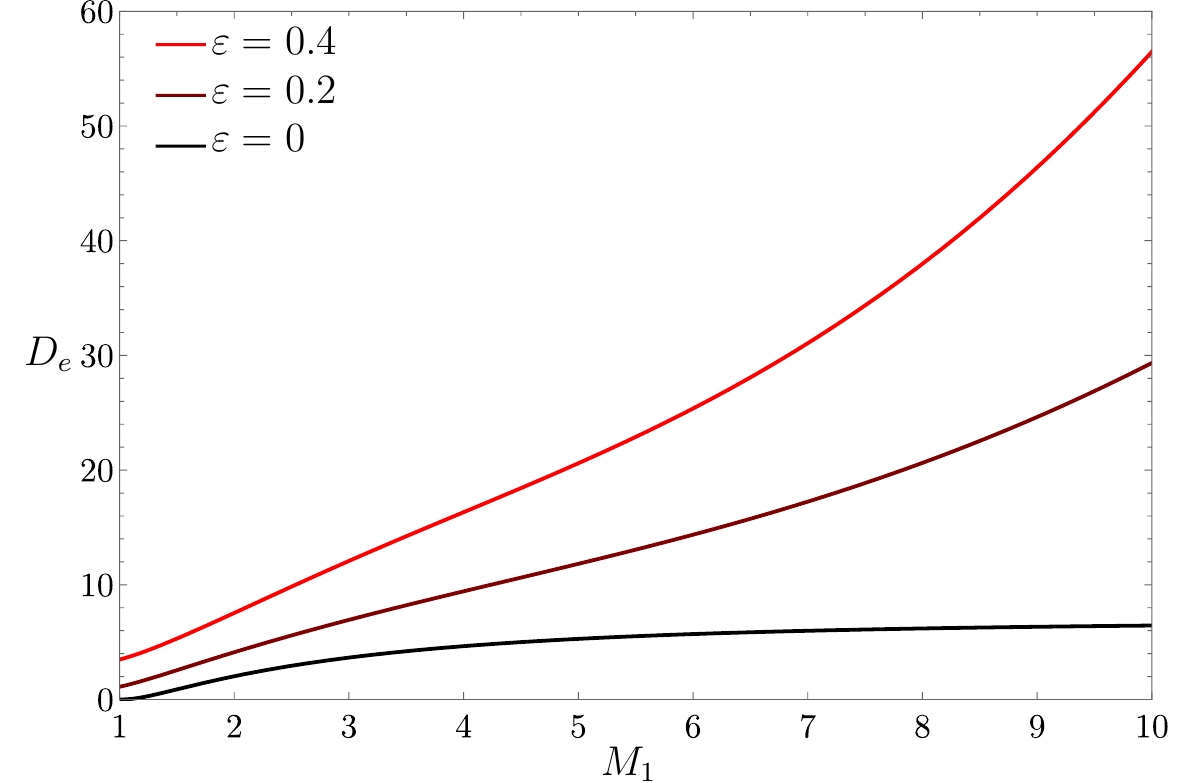} \\
\includegraphics[width=0.47\textwidth]{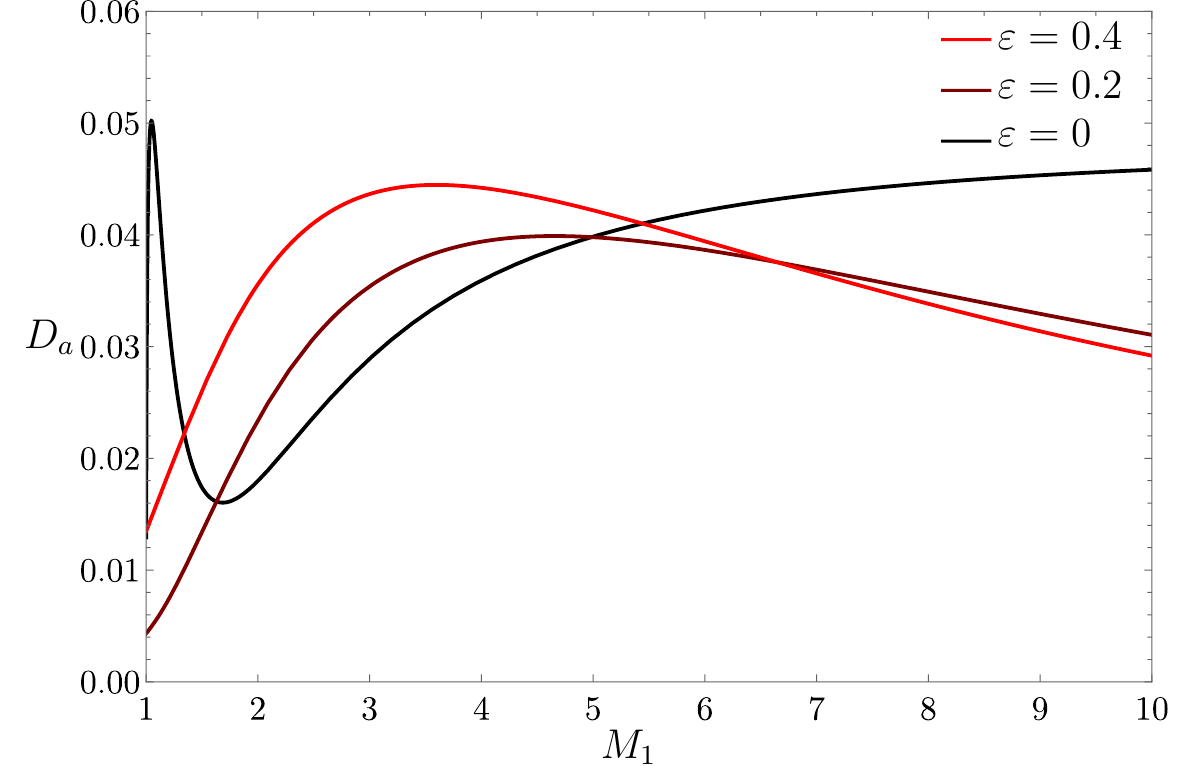}
\caption{Variable $D$ as a function of $M_1$ for $\varepsilon=$ 0, 0.2, and 0.4. The upper panel shows the contribution of entropic-rotational perturbations $D_e$ and the lower panel displays the acoustic contribution $D_a$.} 
\label{fig:D}
\end{figure}

Figure~\ref{fig:D} shows the function $D_e$ and $D_a$ versus $M_1$ for $\varepsilon=0$, $0.2$, and $0.4$. Both $D_e$ and $D_a$ grows with $M_1$ and $\varepsilon$. The acoustic part $D_a$ is at least two orders of magnitude smaller than the entropic part $D_e$ and thus it is negligible. 

\begin{figure}
\includegraphics[width=0.49\textwidth]{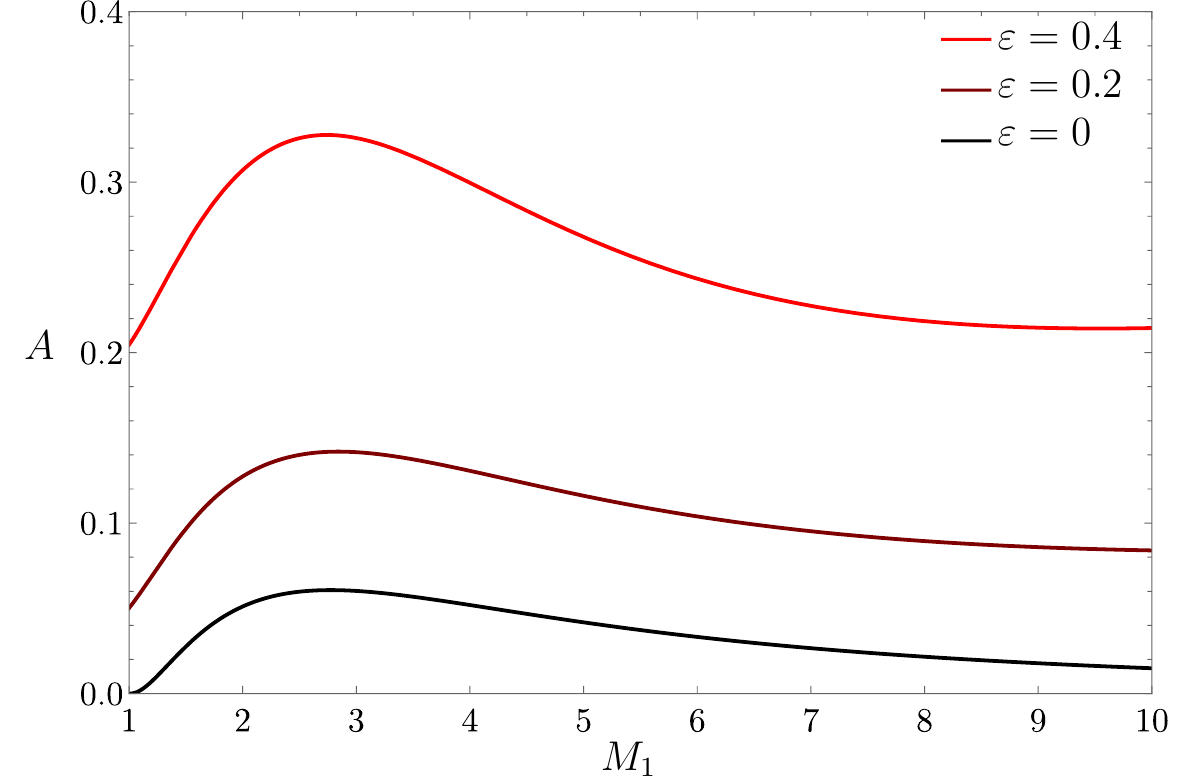}
\caption{Ratio $A=\langle \bar\rho^2_2 \rangle/\langle \delta M_1^2 \rangle$ as a function of the shock strength $M_1$ for $\varepsilon=0$, $0.2$, and $0.4$.} 
\label{fig:drho2dM1}
\end{figure}

In order to obtain a more intuitive insight, it is useful to express $\langle\bar{\rho}^2\rangle$ as a function of the pre-shock turbulent Mach number. The latter is related to the average upstream velocity perturbations as 
\begin{equation}
\begin{aligned}
\langle \delta M_1^2 \rangle=3 \left(\frac{a_2}{a_1}\right)^2\langle \bar{u}_1^2 \rangle=\frac{3 M_1^2}{M_2^2C_2^2}\langle \bar{u}_1^2 \rangle
\end{aligned}
\end{equation}
Combining this with (\ref{uw3D}) and (\ref{denave}), we obtain
\begin{equation}
\begin{aligned}
\langle \bar\rho^2_2 \rangle = \frac{M_2^2C_2^2 D}{8 \pi M^2_1} \langle \delta M_1^2 \rangle = A \langle \delta M_1^2 \rangle
\end{aligned}
\end{equation}
Figure~\ref{fig:drho2dM1} shows the ratio $A=\langle \bar\rho^2_2 \rangle/\langle \delta M_1^2 \rangle$ as a function of the shock strength $M_1$ for $\varepsilon=0$, $0.2$, and $0.4$. For typical values of these parameters ($0.2 \lesssim \varepsilon \lesssim 0.4$ and $M_1 \gtrsim 5$), the ratio $\langle \bar\rho^2_2 \rangle/\langle \delta M_1^2 \rangle$ ranges from $\simeq\!0.1$ to $\simeq\!0.2$. Accordingly, 
\begin{equation}
\label{eq:drho2vsMach1}
\sqrt{\langle \bar\rho^2_2 \rangle} \simeq (0.32 - 0.45) \times \sqrt{\langle \delta M_1^2 \rangle}.
\end{equation}
We can relate the turbulent Mach number $\sqrt{\langle \delta M_1^2 \rangle}$ immediately above the shock to that in the pre-collapse convective shells. During collapse, the Mach number of vorticity waves grows as $\propto r^{(3\gamma-7)/4}$ in the absence of dissipative effects \citep{Kovalenko98,Lai00}. If the convective shell falls from a radius of $\sim 1500\,\mathrm{km}$ to $\sim 200\,\mathrm{km}$, the turbulent Mach number should increase by a factor of $\sim 4.53$. Applying this to scaling (\ref{eq:drho2vsMach1}), we obtain 
\begin{equation}\
\sqrt{\langle \bar\rho^2_2 \rangle} \simeq (1.45 - 2.04) \times \sqrt{\langle \delta M_0^2 \rangle}.
\label{eq:drho2vsMach12}
\end{equation}
The density perturbations predicted by this relation is significantly larger than that generated by the advection of the vorticity waves given by (\ref{eq:drho2_m2}). Below, we investigate if these perturbations contribute to the turbulence in the gain region.

\subsection{Generation of Turbulence from Density Perturbations}

When density perturbations are immersed in a gravitational field, buoyancy effects may play a significant role in contributing to the turbulent kinetic energy. The kinetic energy  production or consumption can be scaled with $\langle \bar{\rho}\bar{u} \rangle g / a_2$ \citep[see, e.g., Chapter 8.2 of][]{Holton12}, with $\bar{u}$ being the velocity component parallel to the gravity field $g$, which in our case coincides with direction of the mean flow. Similarly as done for pure density perturbations, the correlation of velocity and density disturbances can be expressed as
\begin{equation}
\begin{aligned}
\langle\bar{\rho}\bar{u}\rangle = B \int_{0}^\infty\hat{u}_1^2(k)k^2dk
\label{buoave}
\end{aligned}
\end{equation}
where $B$ is a dimensionless pre-spectrum factor,
\begin{equation}
\begin{aligned}
B &= B_{er} + B_a =   \left(\mathcal{D}-1\right)\int_0^{1}\left(\mathcal{U}_{lr}^r\mathcal{P}_{lr}+\mathcal{U}_{li}^r\mathcal{P}_{li}\right)\text{P}(\zeta) {\rm d}\zeta+\\ 
&+\left(\mathcal{D}-1\right)\int_1^{\infty} \mathcal{U}_{s}^r\mathcal{P}_{s}\,\text{P}(\zeta) {\rm d}\zeta + \int_1^{\infty} \mathcal{U}^a\mathcal{P}\, \text{P}(\zeta) {\rm d}\zeta.
\label{B3D}
\end{aligned}
\end{equation}
The entropic-rotational part are the terms proportional to the factor $\mathcal{D}-1$, while the last integral represent the acoustic contribution. For negative values of $\langle \bar{\rho}\bar{u} \rangle$ (i.e., positive velocity-temperature correlation), the density perturbation contributes constructively to the post-shock turbulent kinetic energy. The contrary applies for $\langle \bar{\rho}\bar{u} \rangle>0$.

\begin{figure}
\includegraphics[width=0.47\textwidth]{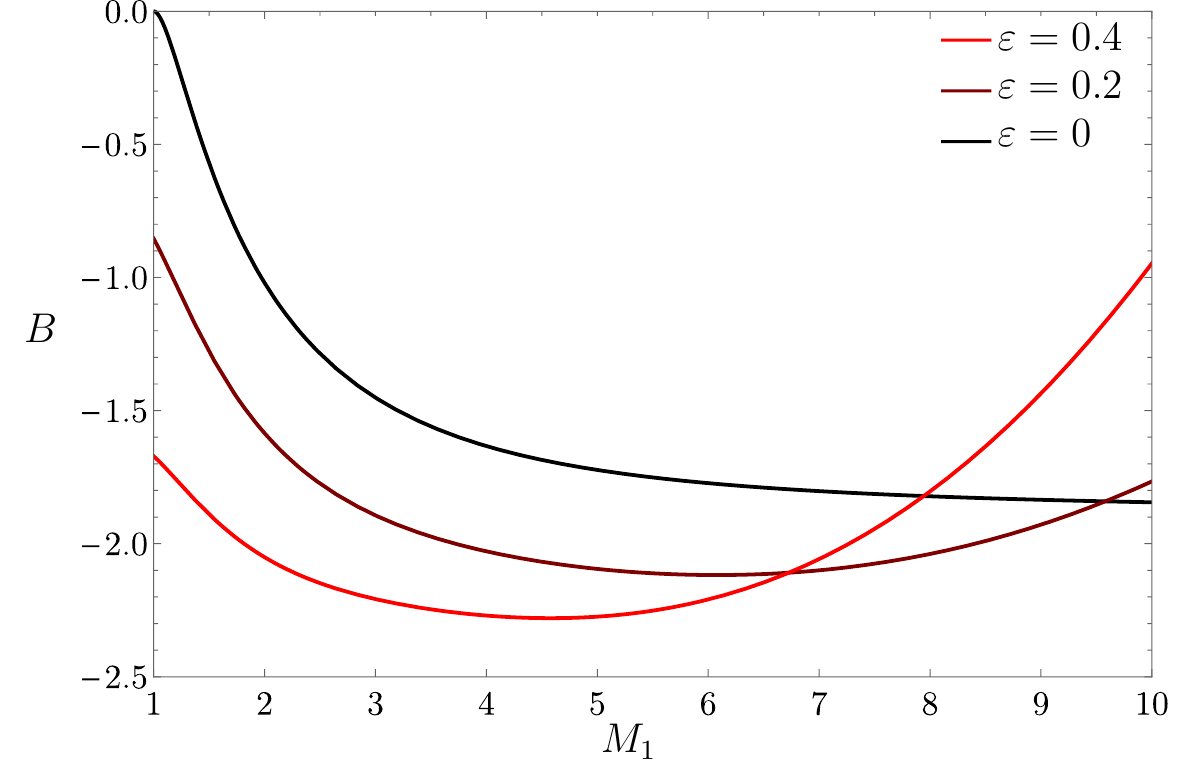}\vspace{2mm}
\caption{Correlated density-velocity $B$ as a function of shock strength $M_1$ for $\varepsilon=$ 0, 0.2, and 0.4.} 
\label{fig:DB}
\end{figure}

Figure \ref{fig:DB} shows $B$ as a function of the the shock Mach number for $\varepsilon=$ 0, 0.2, and 0.4. Similarly to $D$, the acoustic contribution to $B$ is found to be negligible. The buoyancy perturbations are negative, meaning that the density perturbations will increase the value of the final turbulent kinetic energy. 

In the light of this finding, we can substitute the density fluctuations (\ref{eq:drho2vsMach12}) from entropy waves into the expression for the reduction of the critical luminosity (\ref{eq:dl}) and obtain
\begin{equation}
\begin{aligned}
\label{eq:dlf}
\frac{\Delta L_\mathrm{crit}}{L_\mathrm{crit}} \simeq - (1.45 - 2.04) \times  \frac{0.15 \pi}{l \eta_\mathrm{acc}\eta_\mathrm{heat}} \sqrt{\langle \delta M^2_0 \rangle},
\end{aligned}
\end{equation}
For typical values of $\eta_\mathrm{acc}=2$, $\eta_\mathrm{heat}=0.1$, $\sqrt{\langle \delta M^2_0 \rangle} \sim 0.1$, and $l=2$, we get $17-24\%$ reduction in the critical luminosity. This roughly agrees with the results of 3D simulations \citep{Mueller17}. 

\subsection{Impact of acoustic waves and direct injection of kinetic energy}

In addition to the impact of entropy perturbations on the explosion condition of CCNSe, one can in principle study the role of other effects such as the acoustic waves generated by the interaction of the shock with vorticity waves and the direction injection of the kinetic energy of vorticity waves to the post-shock region. The reduction of the critical luminosity due to the latter was estimated by \citet{Abdikamalov16}: 
\begin{equation}
\label{eq:dldi}
\frac{\Delta L_\mathrm{crit}}{L_\mathrm{crit}} \sim 0.6 \langle \delta M^2_1 \rangle .
\end{equation}
For the same parameters used for estimate ($\ref{eq:dlf}$), equation (\ref{eq:dldi}) yields $\sim 12\%$ reduction in the critical luminosity. This is smaller than that due to the entropy perturbations calculated above. Hence, the direct injection of turbulent kinetic energy of vorticity waves is expected to play a sub-dominant effect, in agreement with the estimate of \citet{Mueller15}. 

As we saw above in Sections~\ref{sec:tke} and \ref{sec:density_pert}, the acoustic waves have negligibly small contribution to the perturbations of velocity and density compared to the contributions of the vorticity and entropy modes in the post-shock. For this reason, the acoustic waves in the post-shock region are expected to have negligibly small effect on the explosion condition of CCSNe (see also the discussion in \cite{Mueller16}).  

\section{Conclusions}
\label{sec:conclusion}

The shock-turbulence interplay plays a key role in facilitating core-collapse supernova (CCSN) explosions. In this paper, we studied how vorticity waves from nuclear shell burning affect the shock dynamics once they encounter in the aftermath of stellar core collapse. Our study accounts for the interaction of the shock with intermediate vortical scales, i.e., those whose characteristic length is sufficiently small for the shock to be considered a planar front, yet sufficiently large for the shock to be a seen as a discontinuity front. The mathematical formalism is based on the solution of the linearized hydrodynamics equations in the post-shock region \citep{Wouchuk2009,Huete2017}, which captures the full time evolution of shock-vorticity system in the linear order. In particular, this allowed us to take into account the perturbation of the nuclear dissociation itself, which was not included previously \citep{Abdikamalov16}. We demonstrated that this effect plays an important role in shock-turbulence interaction.

When a vorticity wave encounters a shock, it deforms and generates a post-shock field of vorticity, entropy, and acoustic waves. We have analyzed the properties of these fluctuations for a wide range of the parameters of the incoming vorticity waves and mean flow (Sections~\ref{sec:analysis_wave}-\ref{sec:analysis_field}). We have found that, within the limits of validity of the model, the density perturbations in the post-shock region are dominantly of entropic nature, while the contribution of the acoustic waves is negligibly small. 

We show that the entropy perturbations in the post-shock region is the dominant factor in generating turbulence in the post-shock flow due to work by buoyancy forces (Section~\ref{sec:discussion}). For typical program parameters, the amplitude of density perturbations is about $1.45-2.04$ times the turbulent Mach number in the Si/O shell. Following the method proposed by \cite{Mueller16}, we show that this results in $17-24\%$ reduction in the critical luminosity for producing explosion (cf. Section~\ref{sec:discussion}). This approximately agrees with the results of recent 3D neutrino-hydrodynamics simulations \citep{Mueller17}.  

This paper is the first in a series of two papers that aims at establishing the linear physics of interactions of shocks with turbulent flow. In future, we will study the effect of other perturbation modes that originate from convective shells. Also, the interaction of pre-collapse perturbations with the hydrodynamic instabilities in the post-shock region has to be treated in a more rigorous way as in, e.g., \citet{Takahashi16}. This will be the subject of future studies.

\section*{Acknowledgements}
We thank B. M\"uller for carefully reading the manuscript and for many valuable comments that significantly improved the manuscript. We also thank T. Foglizzo, M. Hempel and A. L. Velikovich for useful discussions. This work is  supported by the Ministry of Science, MEC (ENE2015-65852-C2-2-R) and Fundaci\'on Iberdrola (BINV-ua37crdy), Spain (for C. Huete), by ORAU grant at Nazarbayev University (for E. Abdikamalov), by Max-Planck/Princeton Center (MPPC) for Plasma Physics (NSF PHY-1144374) and a Schmidt Fellowship (for D. Radice). The computations were performed at the NULITS Linux cluster at Nazarbayev University. We thank S. Bubin for his contribution to set up the cluster. 





\begin{thebibliography}{99}
\bibitem[Abdikamalov et al.(2015)]{Abdikamalov15} Abdikamalov, E., Ott, C.~D., Radice, D., et al.\ 2015, \apj, 808, 70 
\bibitem[Abdikamalov et al.(2016)]{Abdikamalov16} Abdikamalov, E., et al.\ 2016, \mnras, 461, 3864--3876 
\bibitem[Adams et al.(2017)]{Adams17} Adams, S.~M., Kochanek, C.~S., Gerke, J.~R., \& Stanek, K.~Z.\ 2017, \mnras, 469, 1445
\bibitem[Audi et al.(2003)]{Audi03} Audi, G., Wapstra, A.~H., \& Thibault, C.\ 2003, Nuclear Physics A, 729, 337 
\bibitem[Baade \& Zwicky(1934)]{Baade34} Baade, W., \& Zwicky, F.\ 1934, Proceedings of the National Academy of Science, 20, 254 
\bibitem[Batchelor(1953)]{Batchelor1953} Batchelor, G.~K. The theory of homogeneous turbulence. Cambridge University Press, 1953.
\bibitem[Bethe(1990)]{Bethe90} Bethe, H.~A.\ 1990, Reviews of Modern Physics, 62, 801
\bibitem[Blondin et al.(2003)]{Blondin03} Blondin, J.~M., Mezzacappa, A., \& DeMarino, C.\ 2003, \apj, 584, 971
\bibitem[Bruenn et al.(2016)]{Bruenn16} Bruenn, S.~W., Lentz, E.~J., Hix, W.~R., et al.\ 2016, \apj, 818, 123 
\bibitem[Burrows \& Goshy(1993)]{Burrows93} Burrows, A., \& Goshy, J.\ 1993, \apjl, 416, L75
\bibitem[Burrows et al.(1995)]{Burrows95} Burrows, A., Hayes, J., \& Fryxell, B.~A.\ 1995, \apj, 450, 830 
\bibitem[Burrows et al.(2007)]{Burrows07} Burrows, A., Dessart, L., Livne, E., Ott, C.~D., \& Murphy, J.\ 2007, \apj, 664, 416
\bibitem[Burrows et al.(2016)]{Burrows16} Burrows, A., Vartanyan, D., Dolence, J.~D., Skinner, M.~A., \& Radice, D.,\ 2017, arXiv:1611.05859
\bibitem[Burrows(2013)]{Burrows13} Burrows, A.\ 2013, Reviews of Modern Physics, 85, 245
\bibitem[Cardall \& Budiardja(2015)]{Cardall15} Cardall, C.~Y., \& Budiardja, R.~D.\ 2015, \apjl, 813, L6 
\bibitem[Chu(1958)]{Chu1958} Chu, B.~T. and Kovasznay,  L.~S.~G., \ 1958, J. Fluids Mech. 3, 494.
\bibitem[Cobos et al.(2014)]{Cobos2014} Cobos-Campos, F. and Wouchuk J.G.,\ 2014 \pre, 90.5,  053007
\bibitem[Collins et al.(2017)]{Collins17} Collins, C., M{\"u}ller, B., \& Heger, A.\ 2017, arXiv:1709.00236 
\bibitem[Couch \& Ott(2013)]{Couch13} Couch, S.~M., \& Ott, C.~D.\ 2013, \apjl, 778, L7
\bibitem[Couch \& Ott(2015)]{Couch15a} Couch, S.~M., \& Ott, C.~D.\ 2015, \apj, 799, 5
\bibitem[Couch et al.(2015)]{Couch15b} Couch, S.~M., Chatzopoulos, E., Arnett, W.~D., \& Timmes, F.~X.\ 2015, \apjl, 808, L21 
\bibitem[Dolence et al.(2013)]{Dolence13} Dolence, J.~C., Burrows, A., Murphy, J.~W., \& Nordhaus, J.\ 2013, \apj, 765, 110
\bibitem[Fern{\'a}ndez et al.(2009a)]{Fernandez2009a} Fernandez, R., et al. \ 2009, \apj, 697, 1827
\bibitem[Fern{\'a}ndez et al.(2009b)]{Fernandez2009b} Fernandez, R., et al. \ 2009, \apj, 703, 1464
\bibitem[Fern{\'a}ndez et al.(2014)]{Fernandez14} Fern{\'a}ndez, R., M{\"u}ller, B., Foglizzo, T., \& Janka, H.-T.\ 2014, \mnras, 440, 2763
\bibitem[Fernandez(2015)]{Fernandez15} Fern{\'a}ndez, R.\ 2015, \mnras, 452, 2071 
\bibitem[Foglizzo \& Tagger(2000)]{Foglizzo00} Foglizzo, T., \& Tagger, M.\ 2000, \aap, 363, 174
\bibitem[Foglizzo et al.(2006)]{Foglizzo06} Foglizzo, T., Scheck, L., \& Janka, H.-T.\ 2006, \apj, 652, 1436
\bibitem[Foglizzo(2009)]{Foglizzo09} Foglizzo, T.\ 2009, \apj, 694, 820 
\bibitem[Foglizzo et al.(2015)]{Foglizzo15} Foglizzo, T., Kazeroni, R., Guilet, J., et al.\ 2015, \pasa, 32, e009
\bibitem[Fraley(1986)]{Fraley1986} Fraley, G., 1986 Phys. Fluids 29, 376--386
\bibitem[Hanke et al.(2012)]{Hanke12} Hanke, F., Marek, A., M{\"u}ller, B., \& Janka, H.-T.\ 2012, \apj, 755, 138
\bibitem[Hanke et al.(2013)]{Hanke13} Hanke, F., M{\"u}ller, B., Wongwathanarat, A., Marek, A., \& Janka, H.-T.\ 2013, \apj, 770, 66
\bibitem[Herant(1995)]{Herant95} Herant, M.\ 1995, \physrep, 256, 117
\bibitem[Huete et al.(2013)]{Huete2013} Huete, C., et al. \ 2013, Phys. Fluids, 25, 076105.
\bibitem[Huete et al.(2017)]{Huete2017} Huete, C., et al.\ sent for publication.
\bibitem[Holton \& Hakim (2012)]{Holton12} Holton, J.~R., Hakim G.~J. 2012, An Introduction to Dynamic Meteorology, 5th edition, Academic Press, Boston.
\bibitem[Jackson(1993)]{Jackson1993} Jackson, T.~L., et al.\ 1993, Phys. Fluids A: Fluid Dyn. 5, 745
\bibitem[Janka \& Mueller(1996)]{Janka96} Janka, H.-T., \& Mueller, E.\ 1996, \aap, 306, 167 
\bibitem[Janka et al.(2012)]{Janka12} Janka, H.-T., Hanke, F., H{\"u}depohl, L., et al.\ 2012, Progress of Theoretical and Experimental Physics, 2012, 01A309
\bibitem[Kashiyama \& Quataert(2015)]{Kashiyama15} Kashiyama, K., \& Quataert, E.\ 2015, \mnras, 451, 2656 
\bibitem[Kovalenko \& Eremin(1998)]{Kovalenko98} Kovalenko, I.~G., \& Eremin, M.~A.\ 1998, \mnras, 298, 861
\bibitem[Kovasznay(1953)]{Kovasznay53} Kovasznay L. S.~G.,  1953, Journal of the Aeronautical Sciences, 20, 657
\bibitem[Lai \& Goldreich(2000)]{Lai00} Lai, D., \& Goldreich, P.\ 2000, \apj, 535, 402
\bibitem[Lattimer \& Swesty(1991)]{Lattimer91} Lattimer, J.~M., \& Douglas Swesty, F.\ 1991, Nuclear Physics A, 535, 331 
\bibitem[Lentz et al.(2015)]{Lentz15} Lentz, E.~J., Bruenn, S.~W., Hix, W.~R., et al.\ 2015, \apjl, 807, L31
\bibitem[Lovegrove \& Woosley(2013)]{Lovegrove13} Lovegrove, E., \& Woosley, S.~E.\ 2013, \apj, 769, 109 
\bibitem[Mabanta et al.(2017)]{Mabanta17} Mabanta, Q. \&  Murphy J.~W.\ 2017, arXiv:1706.00072 
\bibitem[Mahesh et al.(1996)]{Mahesh96} Mahesh K., Moin P.,  Lele S.~K.,  1996, Technical Report TF-69, {The interaction of a shock wave with a turbulent shear flow}. Thermosciences division, Department of Mechanical Engineering, Stanford University
\bibitem[Melson et al.(2015)]{Melson15a} Melson, T., Janka, H.-T., \& Marek, A.\ 2015, \apjl, 801, L24
\bibitem[M{\"o}sta et al.(2014)]{Moesta14} M{\"o}sta, P., Richers, S., Ott, C.~D., et al.\ 2014, \apjl, 785, L29
\bibitem[M{\"o}sta et al.(2015)]{Moesta15} M{\"o}sta, P., Ott, C.~D., Radice, D., et al.\ 2015, \nat, 528, 376
\bibitem[M{\"u}ller \& Janka(2015)]{Mueller15} M{\"u}ller, B., \& Janka, H.-T.\ 2015, \mnras, 448, 2141 
\bibitem[M{\"u}ller et al.(2016)]{Mueller16} M{\"u}ller, B., Viallet, M., Heger, A., \& Janka, H.-T.\ 2016, \apj, 833, 124 
\bibitem[M{\"u}ller(2016)]{Mueller16b} M{\"u}ller, B.\ 2016, \pasa, 33, e048 
\bibitem[M{\"u}ller et al.(2017)]{Mueller17} M{\"u}ller, B., Melson, T., Heger, A., \& Janka, H.-T.\ 2017, arXiv:1705.00620 
\bibitem[Murphy et al.(2013)]{Murphy13} Murphy, J.~W., Dolence, J.~C., \& Burrows, A.\ 2013, \apj, 771, 52
\bibitem[Nadezhin(1980)]{Nadezhin80} Nadezhin, D.~K.\ 1980, \apss, 69, 115 
\bibitem[O'Connor \& Ott(2010)]{Oconnor10} O'Connor, E., \& Ott, C.~D.\ 2010, Classical and Quantum Gravity, 27, 114103 
\bibitem[O'Connor \& Ott(2011)]{Oconnor11} O'Connor, E., \& Ott, C.~D.\ 2011, \apj, 730, 70
\bibitem[Ott et al.(2011)]{Ott11} Ott, C.~D., Reisswig, C., Schnetter, E., et al.\ 2011, Physical Review Letters, 106, 161103
\bibitem[Ott et al.(2013)]{Ott13} Ott, C.~D., Abdikamalov, E., M{\"o}sta, P., et al.\ 2013, \apj, 768, 115 
\bibitem[Ribner(1953)]{Ribner53} Ribner H.~S., 1953, Technical Report {TN 2864}, {Convection of a pattern of vorticity through a shock wave}.
\bibitem[Radice et al.(2015)]{Radice15} Radice, D., Couch, S.~M., \& Ott, C.~D.\ 2015, Computational Astrophysics and Cosmology, 2, 7
\bibitem[Radice et al.(2016)]{Radice16} Radice, D., Ott, C.~D., Abdikamalov, E., et al.\ 2016, \apj, 820, 76 
\bibitem[Radice et al.(2017)]{Radice17} Radice, D., Burrows, A., Vartanyan, D., Skinner, M.~A, Dolence, J.~D., arXiv:1702.03927
\bibitem[Roberts et al.(2016)]{Roberts16} Roberts, L.~F., Ott, C.~D., Haas, R., et al.\ 2016, \apj, 831, 98
\bibitem[Steiner et al.(2013)]{Steiner13} Steiner, A.~W., Hempel, M., \& Fischer, T.\ 2013, \apj, 774, 17 
\bibitem[Summa et al.(2017)]{Summa17} Summa, A., Janka, H.-T., Melson, T., \& Marek, A.\ 2017, arXiv:1708.04154 
\bibitem[Takahashi \& Yamada(2014)]{Takahashi14} Takahashi, K., \& Yamada, S.\ 2014, \apj, 794, 162 
\bibitem[Takahashi et al.(2016)]{Takahashi16} Takahashi, K., Iwakami, W., Yamamoto, Y., \& Yamada, S.\ 2016, \apj, 831, 75 
\bibitem[Takiwaki et al.(2014)]{Takiwaki14} Takiwaki, T., Kotake, K., \& Suwa, Y.\ 2014, \apj, 786, 83
\bibitem[Takiwaki et al.(2016)]{Takiwaki16} Takiwaki, T., Kotake, K., \& Suwa, Y.\ 2016, \mnras, 461, L112
\bibitem[Ugliano et al.(2012)]{Ugliano12} Ugliano, M., Janka, H.-T., Marek, A., \& Arcones, A.\ 2012, \apj, 757, 69
\bibitem[Velikovich et al.(2016)]{Velikovich2016} Velikovich, A.~L., et al.\ 2016, Phys. Plasmas, 23, 052706
\bibitem[Woosley \& Heger(2007)]{Woosley07} Woosley, S.~E., \& Heger, A.\ 2007, \physrep, 442, 269 
\bibitem[Wouchuk et al.(2001)]{Wouchuk2001} Wouchuk, J.~G.\ 2001, \pre, 63.5, 056303
\bibitem[Wouchuk et al.(2001b)]{Wouchuk2001b} Wouchuk, J.~G.\ 2001, Phys. Plasma, 8.6, 2890-2907
\bibitem[Wouchuk et al.(2009)]{Wouchuk2009} Wouchuk, J.~G., et al.\ 2009, \pre, 79, 066315
\bibitem[Yamasaki \& Yamada(2006)]{Yamasaki06} Yamasaki, T., \& Yamada, S.\ 2006, \apj, 650, 291 
\bibitem[Zaidel(1960)]{Zaidel1960} Zaidel, P.~M., J. Appl. Math. Mech. 24, 316.
\end{thebibliography}




\appendix

\section{Laplace transform for the shock temporal evolution and asymptotic expressions} \label{App1}
The transient response of the CCSN shock wave is analyzed employing a similar approach as that used in \citet{Wouchuk2009}, where the transformation
\begin{equation}
x = r \sinh \chi\ , \qquad \tau = r \cosh \chi \label{rchi}
\end{equation}
is conveniently introduced, with $\tau=0$ being the moment at which the initially planar shock front encounters the perturbed field. It is readily seen that $\chi=$const represents a planar front moving in the burnt gas along the $x$ axis, then sweeping the burnt-gas domain from the weak discontinuity $x=0$ $(\chi=0)$ to the detonation front $x=M_2 \tau$ $(\tanh \chi_s=M_2)$. The sound wave equation \eqref{sonicwave} reduces to
\begin{equation}
r\frac{\partial^2 \bar{p}}{\partial r^2} + \frac{\partial \bar{p}}{\partial r} + r \bar{p} = \frac{1}{r}\frac{\partial^2 \bar{p}}{\partial \chi^2}\ ,
\label{sonicrchi}
\end{equation}
while the shock boundary condition at the front becomes
\begin{equation}
\frac{d \xi_s(r)}{d r} = \frac{\sigma_a}{\sqrt{1-M_2^2}} \bar{p}_s(r)+ \frac{\hat{u}_1}{\sqrt{1-M_2^2}}\cos\left(\zeta r\right)\ ,
\label{xisr}
\end{equation}
\begin{equation}
     \begin{aligned}
\left.\frac{1}{r}\frac{\partial \bar{p}_s}{\partial \chi}\right|_s &= -\sigma_b \frac{\partial \bar{p}_s(r)}{\partial r}-\frac{M_2^2 \left(C_2- 1\right)}{\sqrt{1-M_2^2}}\xi_s(r) \\
&+\hat{u}_1 \zeta \frac{C_2-1}{C_2}\sin\left(\zeta r\right)\ ,
\label{psr}
	\end{aligned}
\end{equation}
where $\zeta$ is the characteristic shock oscillation frequency provided in \eqref{zeta}.

The Laplace Transform over the variable $r$ 
\begin{equation}
\mathscr{F}(s,\chi) = \int_0^{\infty}f(r,\chi)e^{-sr}\rm{d} r
\label{Laplace}
\end{equation}
is used to reduce the above shock boundary condition to an algebraical system as function of the Laplace variable $s$. The isolated boundary condition is used to compute the Laplace transform of $\left.\frac{1}{r}\frac{\partial \bar{p}_s}{\partial \chi}\right|_s $ as $\sqrt{s^2+1}\mathscr{P}_s - \bar{p}_{s0}$, thereby giving
\begin{equation}
     \begin{aligned}
\mathscr{P}_s(s) &=\frac{s \left(1+\sigma_b\right)}{s\sqrt{s^2+1}+\sigma_b s^2 + \sigma_c}\bar{p}_{s0} \\ 
& +\frac{s \sigma}{\left(s\sqrt{s^2+1}+\sigma_b s^2 + \sigma_c\right)\left(s^2+\zeta^2\right)}\hat{u}_{1}
	\end{aligned}
    \label{PsLaplace}
\end{equation}
for the Laplace transform of the pressure perturbation at the shock, with
\begin{equation}
\sigma = \frac{C_2-1}{C_2}\left( \zeta^2 -\frac{M_1^2}{M_1^2-1}\right)
\label{sigma}
\end{equation} 
representing the periodic excitation amplitude as a function of $\zeta$, the characteristic oscillation frequency in the $r$-domain variable, and
\begin{equation}
\sigma_c = \frac{M_2^2\left(C_2-1\right)}{1-M_2^2}\sigma_a
\label{Cepsilon}
\end{equation}
 standing for the scaled $\sigma_a$ factor. The solution for the pressure field in \eqref{sonicrchi} can be expressed as a combination of the Bessel functions \citep{Zaidel1960}
\begin{equation}
\bar{p}(r,\chi)=\sum_{\nu=0}^{\infty}A_{\nu} \left( B_{\nu_1}e^{\nu \chi} +B_{\nu_2}e^{-\nu \chi} \right)J_{\nu}(r)
\label{Besselr}
\end{equation}
provided that the solution for Bessel function of the second type must be excluded in order to avoid a divergent behavior when $r\rightarrow 0$. The Laplace transform of \eqref{Besselr} at the shock front yields
\begin{equation}
\mathscr{P}_s(s)=\sum_{\nu=0}^{\infty}N_{\nu} \frac{e^{-\nu \sinh^{-1}s}}{\sqrt{s^2+1}}\ ,
\label{PsBessels}
\end{equation}
with $N_{\nu}= A_{\nu} \left( B_{\nu_1}e^{\nu \chi_s} +B_{\nu_2}e^{-\nu \chi_s} \right)$ referring to the coefficients that accompany the Bessel functions at the shock front in \eqref{psBesseltau}. Combination of \eqref{PsBessels} with \eqref{PsLaplace} is used to provide
\begin{equation}
N_\nu = \frac{N_{\nu-8} a_{4_-}+N_{\nu-6}a_{2_-}+N_{\nu-4} a_s+N_{\nu-2} a_{2_+}}{a_{4_+}}\ 
\label{Dm}
\end{equation}
for the even indices, with the odd values $N_{2\nu+1}$ found all to be zero. The initial coefficients of the recurrence equation are 
\begin{equation}
\begin{aligned}
&N_0 = \frac{1}{a_{4_+}}\ , \\
&N_2 = \frac{D_0 a_{2_+}-b}{a_{4_+}}\ , \\
&N_4 = \frac{D_0 a_s+D_2 a_{2_+}}{a_{4_+}}\ ,\\
&N_6 = \frac{D_0 a_{2_-}+D_2 a_s+D_4 a_{2_+}+b}{a_{4_+}}\ ,\\
&N_8 = \frac{D_0 a_{4_-}+D_2 a_{2_-}+D_4 a_s+D_6 a_{2_+}-1}{a_{4_+}}\ ,
\label{Do8}
\end{aligned}
\end{equation}
with the parameters 
\begin{equation}
\begin{aligned}
&a_{4_+} = -\left(1+\sigma_b\right)\ ,\quad  a_{4_-} = -\left(1-\sigma_b\right)\ , \\
&a_{2_+} = 2\left[2 \sigma_c-\sigma_b+\left(\sigma_b+1\right)\left(2\zeta^2-1\right)\right]\ , \\
&a_{2_-} = 2\left[2 \sigma_c-\sigma_b+\left(\sigma_b-1\right)\left(2\zeta^2-1\right)\right]\ , \\
&a_{4_-} = 2\left[\sigma_b+2\left(2\sigma_b-\sigma_b\right)\left(2\zeta^2-1\right)\right]\ , \\
&b = 2+4\left(\sigma_3-\zeta^2\right)\ .
\label{aa}
\end{aligned}
\end{equation}

The ripple amplitude is easily computed through temporal integration of eq.\eqref{xis} yielding
\begin{equation}
\begin{aligned}
\xi_s(r)&=\sum_{\nu=0}^{\infty}N_{\nu} \frac{r^{k+1}}{2^k \Gamma(2+k)}\mathcal{H}\left(\frac{1+k}{2},\frac{3+k}{2},1+k,-\frac{r^2}{4}\right) \\
&+ \frac{\hat{u}_1}{\zeta\sqrt{1-M_2^2}}\cos(\zeta r)
\label{xisbeseel}
\end{aligned}
\end{equation}
with $\Gamma$ being the Euler gamma function and $\mathcal{H}$ representing the general hyper-geometric function.

In the following, the expressions for the amplitudes of the periodic perturbations downstream are provided. 

The long-time response of the shock pressure evolution is, as shown in \eqref{pstau},
\begin{equation}
\bar{p}_{s}(\tau\gg 1) =  \left \{ \begin{array}{ll}
\mathcal{P}_{lr} \cos\left(\omega_s \tau \right) + \mathcal{P}_{li} \sin\left( \omega_s \tau \right) & ,\zeta \leq 1 \\
\mathcal{P}_{s} \cos\left(\omega_s \tau \right) & ,\zeta \geq 1
	\end{array} \right.
\nonumber
\end{equation}
whose coefficients for the long-wavelength ($\mathcal{P}_{lr}$ and $\mathcal{P}_{li}$) and short-wavelength regime ($\mathcal{P}_{s}$) are:
\begin{equation}
\mathcal{P}_{lr} = \frac{-\sigma \left(\sigma_b\zeta^2-\sigma_c\right)}{\zeta^2 \left(1-\zeta^2\right)+\left(\sigma_b \zeta^2 - \sigma_c\right)}
\label{Plr}
\end{equation}
\begin{equation}
\mathcal{P}_{li} = \frac{\sigma \zeta\sqrt{1-\zeta^2}}{\zeta^2 \left(1-\zeta^2\right)+\left(\sigma_b \zeta^2 - \sigma_c\right)}
\label{Pli}
\end{equation}
\begin{equation}
\mathcal{P}_{s} = \frac{-\sigma }{\zeta\sqrt{1-\zeta^2}+\sigma_b \zeta^2 - \sigma_c}
\label{Ps}
\end{equation}

Similarly, the shock ripple asymptotically oscillates according to
\begin{equation}
\xi_{s}(\tau\gg 1) =  \left \{ \begin{array}{ll}
\mathcal{J}_{lr} \sin\left(\omega_s \tau \right) + \mathcal{J}_{li} \cos\left( \omega_s \tau \right) & ,\zeta \leq 1 \\
\mathcal{J}_{s} \sin\left(\omega_s \tau \right) & ,\zeta \geq 1
	\end{array} \right.
\label{xisasym}
\end{equation}
with the corresponding coefficients being obtained through the shock-pressure variations as
\begin{equation}
\mathcal{J}_{lr} = \frac{\sigma_a}{\omega_s}\mathcal{P}_{lr} + \frac{1}{\omega_s}\, \quad \mathcal{J}_{li} = -\frac{\sigma_a}{\omega_s}\mathcal{P}_{li}
\label{Jlri}
\end{equation}
for the long-wavelength regime, and
\begin{equation}
\mathcal{J}_{s} = \frac{\sigma_a}{\omega_s}\mathcal{P}_{s} + \frac{1}{\omega_s}
\label{Js}
\end{equation}
for the short-wavelength regime.

Only for $\zeta\geq1$ the pressure perturbations escape from the shock in a stable manner. The velocity-acoustic perturbations are given by the long-time irrotational contribution, yielding  
\begin{equation}
\bar{u}_{a}(x,\tau)=\mathcal{U}^a\cos\left(\omega_a\tau-k_a x \right)\ , 
\label{ua}
\end{equation}
and
\begin{equation}
\bar{v}_a(x,\tau)=\mathcal{V}^a\sin\left(\omega_a\tau-k_ax \right)\ ,
\label{va}
\end{equation}
with the 
\begin{equation}
\mathcal{U}^a =\frac{ k_a}{\omega_a}\mathcal{P}_s\ ,\quad \mathcal{V}^a =\frac{1}{\omega_a}\mathcal{P}_s\ ,
\label{UVa}
\end{equation}
being the associated amplitudes.

The asymptotic rotational contribution of the velocity field is 
\begin{equation}
\bar{u}_{r}(x\gg 1) =  \left \{ \begin{array}{ll}
\mathcal{U}_{lr}^r \cos\left( \frac{\omega_s}{M_2} x\right) + \mathcal{U}_{li}^r \sin\left( \frac{\omega_s}{M_2} x\right) & ,\zeta \leq 1 \\
\mathcal{U}_{s}^r \cos\left( \frac{\omega_s}{M_2} x\right) & ,\zeta \geq 1
	\end{array} \right.
\label{urasym}
\end{equation}
and 
\begin{equation}
\bar{v}_{r}(x\gg 1) =  \left \{ \begin{array}{ll}
\mathcal{V}_{lr}^r \sin\left( \frac{\omega_s}{M_2} x\right) + \mathcal{V}_{li}^r \cos\left( \frac{\omega_s}{M_2} x\right) & ,\zeta \leq 1 \\
\mathcal{V}_{s}^r \sin\left( \frac{\omega_s}{M_2} x\right) & ,\zeta \geq 1
	\end{array} \right.
\label{vrasym}
\end{equation}
with the amplitudes $\mathcal{U}^r$ and $\mathcal{V}^r$ being
\begin{equation}
\mathcal{U}_{lr}^r=\frac{\Omega_2 \mathcal{P}_{lr}+\Omega_1}{1+\frac{1-M_2^2}{M_2^2}\zeta^2}\ ,\quad 
\mathcal{U}_{li}^r=\frac{\Omega_2 \mathcal{P}_{li}}{1+\frac{1-M_2^2}{M_2^2}\zeta^2}
\label{Urlri}
\end{equation}
\begin{equation}
\mathcal{U}_s^r=\frac{\Omega_2 \mathcal{P}_s+\Omega_1}{1+\frac{1-M_2^2}{M_2^2}\zeta^2}\ ,
\label{Us}
\end{equation}
for the streamwise component, and 
\begin{equation}
\mathcal{V}_{lr}^r=\frac{\sqrt{1-M_2^2}}{M_2}\zeta\ \mathcal{U}_{lr}^r\ ,\quad 
\mathcal{V}_{li}^r=-\frac{\sqrt{1-M_2^2}}{M_2}\zeta\ \mathcal{U}_{li}^r
\label{Vrlri}
\end{equation}
\begin{equation}
\mathcal{V}_s^r=\frac{\sqrt{1-M_2^2}}{M_2}\zeta\ \mathcal{U}_s^r\ ,
\label{Vs}
\end{equation}
for the crosswise component.

In the absence of pre-shock density perturbations, the amplitude of the density variations downstream is proportional to the shock pressure changes through eq.\eqref{dene}, for the entropic contribution. The acoustic part coincides with the pressure field with the adimensionalization chosen. The total amplitude is then provided by the factor $\mathcal{D}\mathcal{P}$, with $\mathcal{P}$ given in \eqref{Plr}-\eqref{Ps}. 

\begin{figure}
\includegraphics[width=0.47\textwidth]{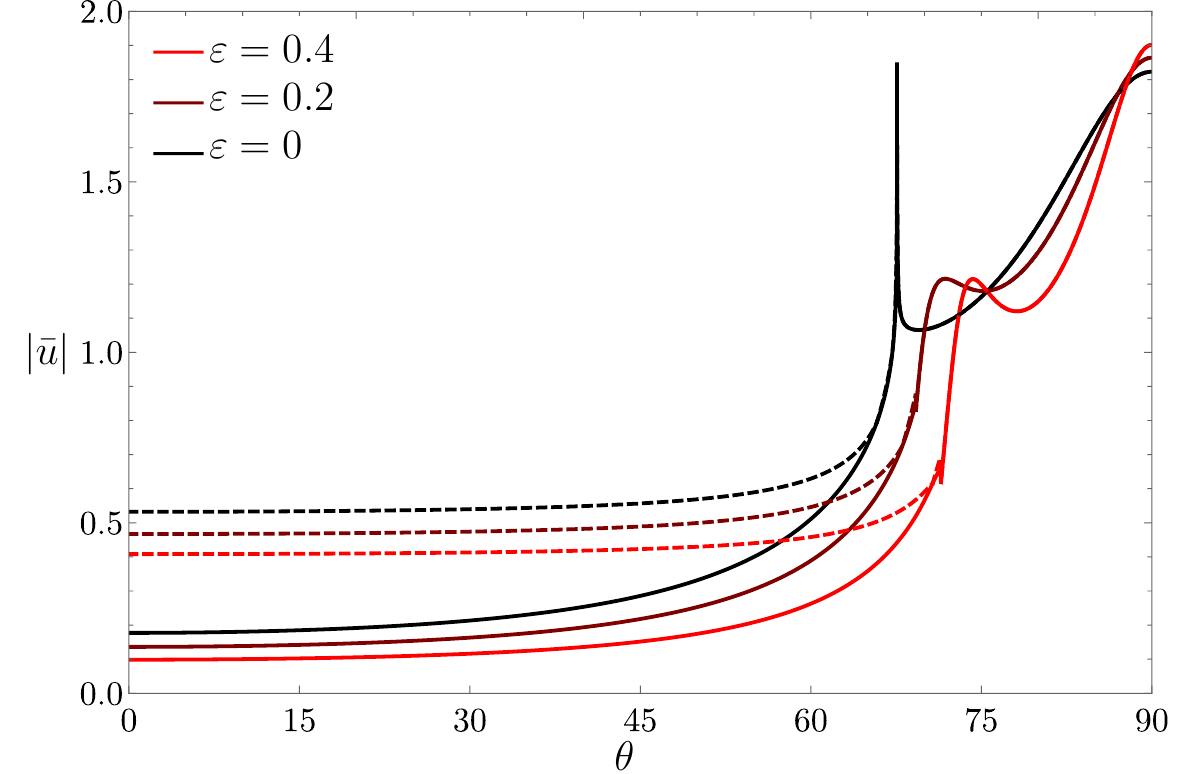} \vspace{2mm}\\  \includegraphics[width=0.47\textwidth]{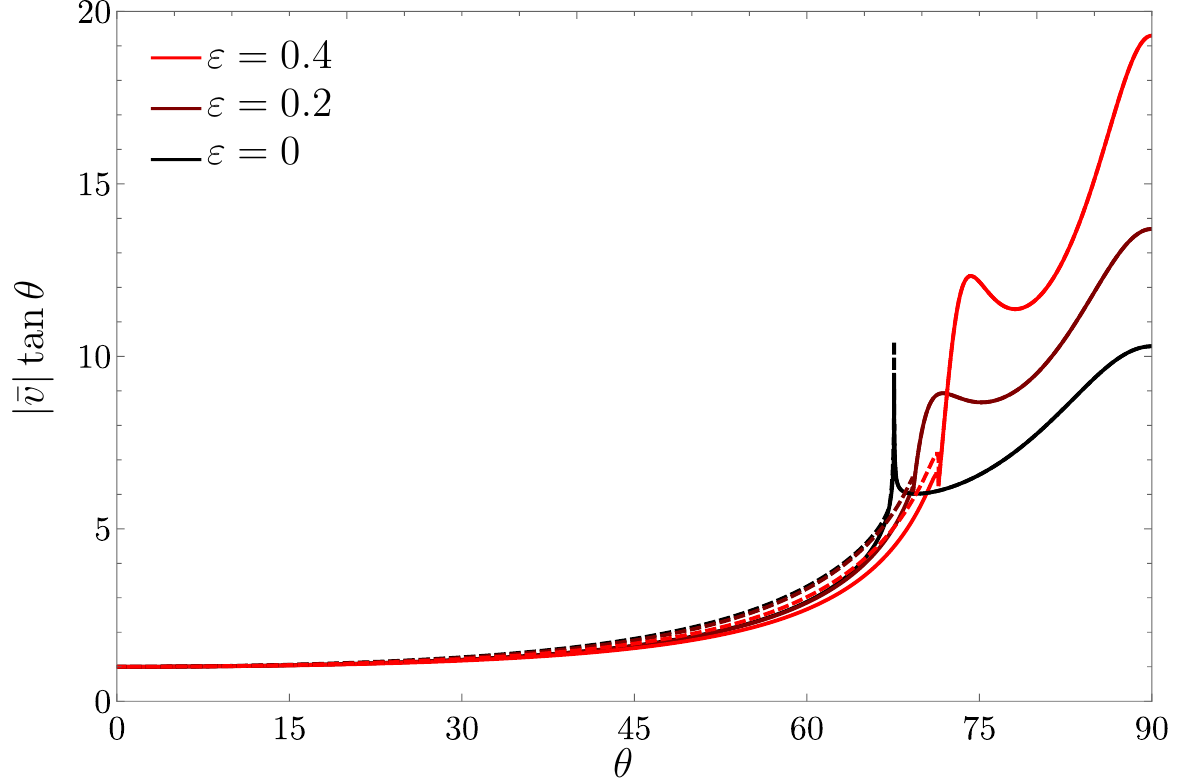}\vspace{2mm}\\
\includegraphics[width=0.47\textwidth]{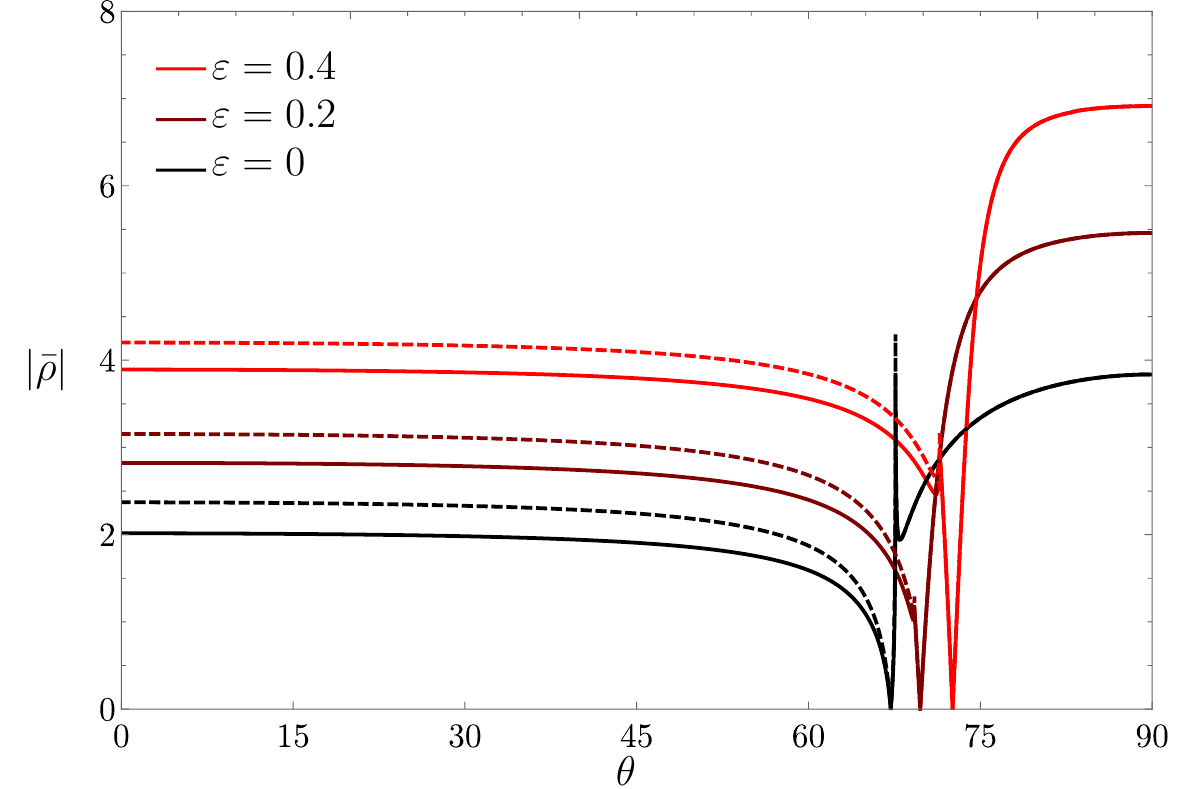}
\caption{Velocity and density perturbation amplitudes for $M_1=5$ and $\varepsilon=$ 0, 0.2, and 0.4. Solid lines show the rotational-entropic contributions, while dashed lines includes the effect of acoustic perturbations.} 
\label{fig:UV}
\end{figure}

The asymptotic value of the nondimensional velocity and density amplitudes, as a function of the incident shear angle $\theta$, are shown in Fig.~\ref{fig:UV} for $\gamma=4/3$, $M_1=5$ and $\varepsilon=$ 0, 0.2, and 0.4. Solid lines show the rotational-entropic contribution and dashed lines include the acoustic contribution. Computations show that each of the amplitudes develops a pronounced finite-amplitude peak at the critical angle $\theta_{cr}$ corresponding to $\zeta=1$. The acoustic contribution is only important for low-incident angles (high-frequency disturbances) in the longitudinal component. The effect of dissociation energy in the peak is to shift the position to higher incident angles, and to make it less pronounced. The former, also found in \citep{Abdikamalov16}, is the result of the change induced by endothermic process in the background properties. The trend  obviously reverses when studying exothermic reactive waves as detonations \citep{Huete2017}. The latter effect is however result of the perturbation of the energy employed in dissociating nuclei.

\section{Nuclear dissociation parameter}
\label{App2}

The nuclear dissociation at the shock can be parametrized in terms of the local free-fall velocity \citep{Fernandez2009a,Fernandez2009b}:
\begin{equation}
\label{eq:eps}
\Delta e_\mathrm{dis} = \frac{1}{2} \varepsilon \upsilon_\mathrm{FF}^2,
\end{equation}
where $\varepsilon$ is a dimensionless dissociation parameter. This parameter is related to the change of nuclear mass fractions in the following way. Consider nuclei with an atomic mass number $A_i$ in a fluid element of mass $M$ that contains multiple types of nuclei. Hereafter, index $i$ is used to refer to these nuclei. Suppose that these nuclei have a mass fraction of $X_i$ in the fluid element, meaning that the total mass of these nuclei in the fluid element is 
\begin{equation}
  M_i = X_i M
\end{equation}
Thee total number $N$ of these nuclei in the fluid element is then 
\begin{equation}
N = \frac{M_i}{A_i m_i},
\end{equation}
where $m_u$ is the atomic mass unit. Suppose $e_i$ is the binding energy per nucleon for these nuclei. Thus, the binding energy per nucleus is $A e_i$. The total binding energy in the fluid element is simply $NAe_i$, from which one can easily obtain the binding energy per mass
\begin{equation}
E_i = \frac{NAe_i}{M} = \frac{M_i}{A_i m_u} \frac{A e_i}{M} = \frac{M_i}{M} \frac{e_i}{m_u} = X_i \frac{e_i}{m_u}.
\end{equation}
Thus, if the mass fraction of these nuclei changes by $\Delta X_i$ at shock crossing, the total binding energy per unit mass changes by
\begin{equation}
\label{eq:e_i}
\Delta E_i = \Delta X_i \frac{e_i}{m_u}.
\end{equation}
If $X_i<0$, i.e. the fraction of these nuclei decreases, then $\Delta E_i < 0 $, i.e. the total binding energy of the nuclei in the fluid decreases. This means that the fluid absorbs energy $ - \Delta E_i$ for the dissociation per unit mass. The total dissociation energy at the shock is equal to the sum of $\Delta E_i$ given by equation \eqref{eq:e_i} for all the nuclei that exists in the fluid element:
\begin{equation}
\Delta e_\mathrm{dis} = \sum_i \Delta E_i 
\end{equation} 
Thus, using equation \eqref{eq:eps}, the nuclear dissociation parameter can be calculated as 
\begin{equation} 
\label{eq:varepsilon2}
\varepsilon = \frac{2 \Delta e_\mathrm{dis} }{m_u \upsilon_\mathrm{FF}^2}.
\end{equation} 
In this work, the mass fractions are measured within $2-3$ grid points above and below the shock. Extensive tests with different numbers of grid points have been performed in order to establish the validity of this approach.


\bsp	
\label{lastpage}
\end{document}